\documentclass[aps,pre,reprint,superscriptaddress]{revtex4-1}

\usepackage{amssymb,amsmath,bm}
\usepackage{graphicx}
\usepackage{enumerate}
\usepackage[utf8]{inputenc}
\usepackage{cancel}
\usepackage{color}
\usepackage{hyperref}
\usepackage{float}

\newcommand{\eq}{\text{eq}}
\newcommand{\ini}{\text{i}}
\newcommand{\fin}{\text{f}}

\newcommand{\dorc}{d \text{ or } c}
\newcommand{\kin}{\text{kin}}
\newcommand{\qs}{\text{qs}}

\newcommand{\calS}{\mathcal{S}}
\newcommand{\calL}{\mathcal{L}}
\newcommand{\calP}{\mathcal{P}}

\newcommand{\pder}[2]{\frac{\partial #1}{\partial #2}}
\newcommand{\mean}[1]{\overline{#1}}

\begin{document}


\title{Finite-time adiabatic processes: derivation and speed limit}



\author{Carlos A. Plata}
\affiliation{Dipartimento di Fisica e Astronomia ``Galileo Galilei'',
  Istituto Nazionale di Fisica Nucleare, Università di Padova, Via Marzolo 8, 35131 Padova, Italy}
\author{David Guéry-Odelin}
\affiliation{Laboratoire de Collisions Agrégats Réactivité, CNRS, UMR 5589, IRSAMC, France}
\author{Emmanuel Trizac}
\affiliation{LPTMS, CNRS, Université Paris-Sud, Université Paris-Saclay, 91405 Orsay,
France}
\author{Antonio Prados}
\affiliation{Física Teórica, Universidad de Sevilla, Apartado de
  Correos 1065, E-41080 Sevilla, Spain}


\date{\today}

\begin{abstract}
  Obtaining adiabatic processes that connect equilibrium states in a
  given time represents a  challenge for mesoscopic systems. In
  this paper, we explicitly show how to build these finite-time
  adiabatic processes for an overdamped Brownian particle in an
  arbitrary potential, a system that is relevant both at the
  conceptual and the practical level. This is achieved by jointly
  engineering the time evolutions of the binding potential and the
  fluid temperature. Moreover, we prove that the second principle
  imposes a speed limit for such adiabatic transformations: there
  appears a minimum time to connect the initial and final states. This
  minimum time can be explicitly calculated for a general
  compression/decompression situation.

\end{abstract}


\maketitle


\section{Introduction}\label{sec:intro}

Adiabatic processes are a cornerstone in the thermodynamics of
macroscopic systems. Therein, energy is solely exchanged as
work--there is no heat--and the large system size makes
fluctuations mostly irrelevant. If, in addition, the system always sweeps
equilibrium states, i.e., the process is reversible, there is no
entropy change. These processes played a central conceptual role in laying the foundations
of thermodynamics, culminating with the works of Carath\'eodory and Planck \cite{callen_thermodynamics_1985}. Besides, such processes
are essential to build thermal engines. The Carnot cycle indeed
consists of two reversible isothermal and two reversible
adiabatic branches~\cite{callen_thermodynamics_1985}~\footnote{We
    understand adiabatic in the thermodynamical sense, not in
    the often found ``slow enough'' quantum mechanical meaning, to
    which we refer as {\em quasi-static}.}.

The relevance of mesoscopic systems spreads out across a wide range of
fields in physics and technology, such as
nanodevices~\cite{jin_non-equilibrium_2010,roche_harvesting_2015},
biomolecules~\cite{collin_verification_2005,prados_sawtooth_2013,lin_theories_2018}
or active
matter~\cite{bertin_mesoscopic_2013,battle_broken_2016}. Statistical
methods are typically applicable to mesoscopic systems, but their
smallness entails that fluctuations play an important
role~\cite{sekimoto_stochastic_2010,seifert_stochastic_2012}. It is
thus challenging but also compelling to extend macroscopic concepts to
the mesoscale because new physics often emerges, like the fluctuation
theorems or transient violations of the second
law~\cite{gallavotti_dynamical_1995,jarzynski_nonequilibrium_1997,crooks_nonequilibrium_1998,ritort_work_2004,marconi_fluctuationdissipation:_2008}.

At the mesoscale, defining and characterising adiabatic processes is
crucial, e.g. to build a mesoscopic version of the Carnot engine. But
this is far from trivial: it is meaningless to imagine an {inherently
  fluctuating} Brownian particle thermally isolated from its
environment for each of its trajectories: over them, both work and
heat contribute to the energy
change~\cite{crooks_work_2007,martinez_brownian_2016}.  However, one
can think of processes in which the average heat vanishes, not only
between the initial and final states, but along the whole dynamics;
thus the average work yields the average energy increment. This is the
concept of adiabatic process that we employ throughout.

Finite-time adiabatic processes have not been devised so far. In fact, adiabatic processes have been only analysed in simple limiting cases: vanishing or infinite time operation.  In the overdamped regime, \textit{instantaneous} processes in which the position distribution does not change have been termed adiabatic~\cite{schmiedl_efficiency_2008,holubec_exactly_2014,rana_single-particle_2014,singh_low-dissipation_2018} because the configurational contribution to the heat
vanishes. However, these processes are not actually adiabatic, since
there is a contribution to the heat--and to the entropy change--coming
from the velocity degree of freedom: the temperature varies in such
instantaneous processes~\cite{hondou_unattainability_2000,schmiedl_efficiency_2008}.
For underdamped dynamics, only quasi-static \textit{reversible adiabatic processes} have been analysed, mainly for the harmonic case. Therein, this has led to the condition $T^{2}/k=\text{const}$, where $T$ is the bath temperature and $k$ is the stiffness of the trap \cite{bo_entropic_2013,martinez_adiabatic_2015,martinez_brownian_2016}.

Engineering adiabatic processes requires the joint time control of
both the bath temperature and the confining potential, which can be
implemented in experiments with micron-size colloids manipulated by
laser tweezers in a suspending fluid
\cite{martinez_effective_2013,ciliberto_experiments_2017}.
Optical confinement makes it possible to control the time
dependence of the effective temperature seen by the Brownian
particle~\cite{martinez_effective_2013}. The
dynamics is neatly overdamped for the broad class of systems consisting of mesoscopic
objects suspended in a solvent~\footnote{The time variation of the physical properties
controlling the dynamics--trap stiffness and bath temperature--must be much slower than the
relaxation of velocity to equilibrium, governed by
the fluid viscosity.}. We shall thus carry our analysis in this
framework, see Appendix \ref{sec:overdamped}.

Hereafter, we answer two physically relevant questions. First, we show
how \textit{finite-time adiabatic} processes can be 
built for a colloidal particle driven by an arbitrary potential. 
Not only does this have theoretical
importance but also practical consequences. For example, shortening
the duration of the adiabatic branches of a Brownian Carnot
engine--like the one investigated in
Ref.~~\cite{martinez_brownian_2016}--increases the delivered power.
Second, we show that there appears a fundamental speed limit for such
adiabatic processes. This is at variance with the isothermal case,
where equilibration can be arbitrarily accelerated
\cite{martinez_engineered_2016}.  The emergence of such speed limits
is also of fundamental
interest, with relevance in control theory and the foundations of non-equilibrium statistical mechanics~\cite{sekimoto_complementarity_1997,aurell_refined_2012,ito_stochastic_2018,okuyama_quantum_2018,shiraishi_speed_2018,ito_stochastic_2018,rosales-cabara_optimal_2020}.

The paper is organised as follows. In Sec.~ \ref{sec:engineering}, we
rigorously show the feasibility of finite-time adiabatic processes in
the context of stochastic thermodynamics. Section
\ref{sec:comp-decomp} is devoted to the optimisation of such
processes, either minimising the connecting time or optimising the
target temperature. Finally, we summarise the conclusions of this work
in Sec.~\ref{sec:conclusion}, along with a discussion of future
perspectives. The appendices deal with some technical aspects and
complementary discussions that are not essential for the understanding
of our results, and thus are omitted in the main text.

\section{Engineering finite-time adiabatic processes and speed limit}
\label{sec:engineering}

We consider a Brownian particle immersed in a heat bath at temperature
$T(t)$ and trapped in a generic potential $U(X,t)$.  The particle
position obeys the Langevin equation 
\begin{equation}
\lambda\, dX/dt=-\partial_X
U(X,t)+\sqrt{2\lambda k_{B}T(t)}\xi(t),    
\end{equation}
with $\lambda$ the friction
coefficient and $\xi(t)$ a unit-variance Gaussian white noise. The
Fokker-Planck (FP) equation for the probability density function (PDF)
$P(X,t)$ of finding the particle at position $X$ at time $t$ thus
reads
\begin{equation}\label{eq:Fokker-Planck-with-dimensions}
 \lambda \partial_t P(X,t)=\partial_X \left[\partial_X U(X,t)\, 
    P(X,t)\right]+k_{B}T(t)\partial_X^2 P(X,t).
\end{equation}

We are interested in processes that connect two given equilibrium states in a running time $t_{\fin}$. Dimensionless variables are introduced with the definitions $\tau=t/t_{\fin}$ ($0\leq\tau\leq 1$), $x=X/\sigma_{X,\ini}$, $\theta=T/T_{\ini}$,   $u=U/(k_{B}T_{\ini})$, $p(x,\tau)=\sigma_{X,\ini}\,P(\sigma_{X,\ini}\,x,t_{\fin}\,\tau)$. For any physical quantity $Y$, we denote throughout the paper derivatives by $\dot{Y} \equiv\partial_{\tau}Y$ and $Y'\equiv\partial_{x}Y$, the initial (final) value by $Y_{\ini}$ ($Y_{\fin}$), the difference between final and initial values by $\Delta Y\equiv Y_{\fin}-Y_{\ini}$, and the variance by $\sigma_{Y}^{2}$. The FP equation is then
\begin{subequations}\label{eq:Fokker-Planck}
\begin{align}
  \label{eq:F-P-dimless}
  \dot{p}(x,\tau)=&-j'(x,\tau), \\
  \label{eq:J-dimless}
  \frac{j(x,\tau)}{t_{\fin}^{*}}=&-\left[u'(x,\tau)
              p(x,t)+\theta(\tau)p'(x,\tau)\right],
\end{align}
\end{subequations}
where
$t_{\fin}^{*}=k_{B}T_{\ini}t_{\fin}/(\lambda \sigma_{X,\ini}^{2})$ is
the dimensionless connecting time~\footnote{We
drop the asterisk not to clutter our formulas.}.

Energy has two contributions: a kinetic one, and a configurational one
coming from the potential $u(x,\tau)$. Within the overdamped
description, the average kinetic energy always has the equilibrium
value $\theta/2$, which is time-dependent. Thus, the average energy is
$\overline{E}=\theta/2+\overline u(x,\tau)$, where
$\overline u(x,\tau)=\int dx\, u(x,\tau)p(x,\tau)$.  Work and heat
exchange rates are $\dot{\overline W}=\int dx\, \dot{u}\, p$ and
$\dot{\overline Q}=\dot\theta/2+\dot{\overline Q}_{x}$, with
$\dot{\overline Q}_{x}\equiv\int dx\, u\, \dot{p}=\int dx\, u'\,j$ the
configurational heat rate. The first principle then holds:
$\dot{\overline E}=\dot{\overline Q}+\dot{\overline
  W}$~\cite{sekimoto_stochastic_2010}.

Entropy is introduced as \cite{sekimoto_stochastic_2010}:
$S=S_{\kin}+S_{x}$, where $S_{\kin}=\frac{1}{2}\ln\theta$,
$S_{x}(\tau)=-\int dx\, p(x,\tau)\ln\! p(x,\tau)+K$. We choose the
constant $K$ to make $S_{x,\ini}=0$ and simplify some formulas.  From
the FP equation, extended forms of the second principle have been
derived, $\dot{S}=\dot{S}_{\text{irr}}+\dot{\overline Q}/\theta$,
where $\dot{S}_{\text{irr}}\geq 0$ is the entropy production
rate~\cite{ge_extended_2009,sekimoto_stochastic_2010,seifert_stochastic_2012}. For
adiabatic processes, $\dot{S}_{\text{irr}}$ only contributes to the
entropy change and one gets in dimensionless variables
\begin{equation}\label{eq:second-principle}
  \dot{S} \,= \, \dot{S}_{\text{irr}} \, = \, \frac{1}{t_{\fin}} \, \frac{1}{\theta(\tau)}\int
  dx\,
  \frac{j^{2}(x,\tau)}{p(x,\tau)}.
\end{equation}

Let us consider given equilibrium initial and final states,
corresponding to temperature and potential pairs
$(\theta_{\ini},u_{\ini}(x))$ and $(\theta_{\fin},u_{\fin}(x))$,
respectively. Our first aim is to show the feasibility of connecting
these states adiabatically, i.e. find solutions of
Eq.~\eqref{eq:Fokker-Planck} that (i) have the  canonical
form at both the initial and final times,
\begin{equation}\label{eq:bc-time}
p_{\ini}(x)= Z_{\ini}^{-1}e^{-u_{\ini}(x)/\theta_{\ini}}, \quad 
p_{\fin}(x)=Z_{\fin}^{-1}e^{-u_{\fin}(x)/\theta_{\fin}},
\end{equation}
with $Z_{\ini,\fin}$ ensuring the normalisation of the distributions, and (ii) make the
total heat exchange rate $\dot{\overline Q}=0$ for all times. We show
below how this can be done by tuning the temperature $\theta(\tau)$
and the potential $u(x,\tau)$. Note that $\overline W=\Delta \overline{E}$, regardless of the process duration.

We build explicitly these adiabatic processes by an
\textit{inverse-engineering} procedure. Starting from any 
$p(x,\tau)$ connecting these two fixed
states, Eq.~\eqref{eq:F-P-dimless} gives
$j(x,\tau)=\int_{x}^{+\infty} d\xi\, \dot{p}(\xi,\tau)$.
If we knew $\theta(\tau)$ (we do not yet), integration of
Eq.~\eqref{eq:second-principle} from $\tau=0$ to $1$ would yield
the value of $t_{\fin}$ and Eq.~\eqref{eq:J-dimless} would finally
give the force field $u'(x,\tau)$.
This remark suggests to get rid of $\theta(\tau)$ by
introducing the change of variable $\Xi(\tau)=e^{2S(\tau)}$, with
$\Xi_{\ini}=1$. Then, $\Xi$ evolves according to
\begin{equation}\label{eq:second-principle-rewritten}
  \Xi(\tau)=\theta(\tau)e^{2S_{x}(\tau)}=
  1+\frac{2}{t_{\fin}}\int_{0}^{\tau} d\zeta\,
  e^{2S_{x}(\zeta)} \int
  dx\, \frac{j^{2}(x,\zeta)}{p(x,\zeta)}.
\end{equation}
Starting again from a PDF $p(x,\tau)$ verifying
Eq.~\eqref{eq:bc-time}, $j(x,\tau)$ and also $S_{x}(\tau)$
follow. Thus, we know $\Xi(\tau)$ for all times and we can complete
the inverse-engineering procedure: (i) particularising
Eq.~\eqref{eq:second-principle-rewritten} for $\tau=\tau_{\fin}=1$, we
obtain the value of the connecting time
\begin{equation}\label{eq:delta-value}
  t_{\fin}=\frac{2}{\Delta\Xi}\int_{0}^{1}d\tau\,
  e^{2S_{x}(\tau)} \int dx\, \frac{j^{2}(x,\tau)}{p(x,\tau)},
\end{equation}
 (ii) turning to
Eq.~\eqref{eq:second-principle-rewritten}, we get the temperature program
$\theta(\tau)$--$\theta(\tau)>0$ for all times--and (iii)
Eq.~\eqref{eq:J-dimless} provides us with the force $u'(x,\tau)$ that
does the job.

Equation~\ref{eq:second-principle-rewritten} shows that two
arbitrary states cannot be connected with an adiabatic
transformation. The positiveness of the right hand side ensures that
$\Delta\Xi\geq 0$ or $\Xi_{\fin}\geq \Xi^{\qs}=1$.  The equality only
holds for the quasi-static case: if $\Delta\Xi=0$, we have that
$t_{\fin}$ diverges~\footnote{The current associated to the
  quasi-static solution
  $p^{\qs}(x,\tau)\propto\exp\left[-u(x,\tau)/\theta(\tau)\right]$ does
  not vanish in general,
  $j^{\qs}(x,\tau)=\int_{x}^{\infty}d\xi \dot p^{\qs}(\xi,\tau)\neq
  0$. Then, Eq.~\eqref{eq:J-dimless} poses no problem because it
  becomes an identity.} and $\Xi(\tau)=1$.  Note that, with the
exception of the quasi-static case, the adiabatic process cannot be
reversed in time because that would violate the second principle.

Moreover, the second principle imposes a speed limit for finite-time
adiabatic processes: there appears a minimum non-vanishing value for
the connecting time $t_{\fin}$, except for a trivial
``configurationally-static'' case.  Starting from
Eq.~\eqref{eq:second-principle-rewritten}, this can be proved by a
\textit{reductio ad absurdum} argument. Let us assume that there is no
lower bound for $t_{\fin}$ and thus an instantaneous adiabatic process
with $t_{\fin}=0$ is possible. The rhs of Eq.~\eqref{eq:delta-value}
then vanishes and $j(x,\tau)=0$ everywhere.  This entails that the FP
equation must have a time-independent solution $p(x,\cancel{\tau})$,
but $p_{\ini}(x)\neq p_{\fin}(x)$ in general. This contradiction
completes the argument. If $p_{\ini}(x)= p_{\fin}(x)$,
i.e. $u'_{\fin}(x)/\theta_{\fin}=u'_{\ini}(x)/\theta_{\ini}$ as
follows from Eq.~\eqref{eq:bc-time}, we deal with a
``configurationally-static'' situation and $t_{\fin}$ may vanish--see
below. In that case, the system cannot cool since $\Xi(\tau)$ is
non-decreasing, and thus so is $\theta(\tau)$.

\section{Compression/decompression processes: Optimal connection}
\label{sec:comp-decomp}

Let us analyse a generic and physically relevant case: the compression
or decompression of a system around a fixed average value
$\overline{x}$ (the axis origin for convenience).  We take
$p(x,\tau)=(Z\sigma_{x}(\tau))^{-1}\exp[-u_{\ini}(x/\sigma_{x}(\tau))]$,
thus of shape-preserved form, where $\sigma_{x}$ is the variance of
the distribution and the normalisation constant
$Z=\int dy \exp[-u_{\ini}(y)]$ does not depend on $\sigma_{x}$. The
system is being decompressed (compressed) for $\dot{\sigma}_{x}>0$
($\dot{\sigma}_{x}<0$)~\footnote{All the central moments are
  $\overline{x^{n}(\tau)}=
  (\overline{x^{n}})_{\ini}[\sigma_{x}(\tau)]^{n}$;
  $\sigma_{x,\ini}=1$ and $\theta_{\ini}=1$ with our choice of
  units.}.  The corresponding current and entropy follow immediately
as $j=(\dot{\sigma}_{x}/\sigma_{x})xp$ and
$\Xi=\theta \sigma_{x}^{2}$.  The adiabatic inequality simplifies to
$\theta_{\fin}\sigma_{x,\fin}^{2}\geq 1$.

For each choice of the function $\sigma_{x}(\tau)$ obeying
$\sigma_{x}(0)=1$, $\sigma_{x}(1)=\sigma_{x,\fin}$, the initial and
final states are connected. Herein, explicit expressions for the
connecting time $t_{\fin}$, the temperature program $\theta(\tau)$ and
the binding potential $u(x,\tau)$ can be given. Indeed,
Eqs.~\ref{eq:second-principle-rewritten} and \ref{eq:delta-value}
reduce to
\begin{subequations}\label{eq:theta-tauf-comp-decomp}
\begin{equation}\label{eq:second-principle-particular-case}
  \theta(\tau)\sigma_{x}^{2}(\tau)-1=
  \frac{1}{2t_{\fin}}
  \int_{0}^{\tau}d\zeta \,\left[\frac{d}{d\zeta}
    \sigma_{x}^{2}(\zeta)\right]^{2} ,
\end{equation}
\begin{equation}\label{eq:tf-particular-case}
  t_{\fin}=
  \frac{J[\sigma_{x}]}{2\Delta(\theta \sigma_{x}^{2})}
  \quad\hbox{with}\quad
  J[\sigma_{x}]\equiv\int_{0}^{1}d\tau \,\left[\frac{d}{d\tau}
    \sigma_{x}^{2}(\tau)\right]^{2} 
\end{equation}
\end{subequations}
and the potential stems from Eq.~\eqref{eq:J-dimless},
\begin{equation}\label{eq:potential-adiab-job}
  u(x,\tau)=-\frac{1}{2t_{\fin}}\frac{\dot
    \sigma_{x}(\tau)}{\sigma_{x}(\tau)}x^{2}+\theta(\tau)\,
  u_{\ini}\!\left(\frac{x}{\sigma_{x}(\tau)}\right).
\end{equation}
The speed limit for the adiabatic process can be explicitly
worked out as well. Since the denominator of $t_{\fin}$ in
Eq.~\eqref{eq:tf-particular-case} is fixed, the minimum time
$\tilde{t}_{\fin}$ corresponds to the variance profile
$\tilde{\sigma}_{x}$ that minimises $J[\sigma_{x}]$.  We get
\begin{equation}\label{eq:sigma2-evol-adiab}
  \tilde{\sigma}_{x}(\tau)=\sqrt{1+\tau\, \Delta(\sigma_{x}^{2})}, 
  \quad \hbox{and} \quad
  \tilde{t_{\fin}}=\dfrac{\left(\Delta(\sigma_{x}^{2})\right)^{2 }}{2\,\Delta(\theta \sigma_{x}^{2})}.
\end{equation}
Note that $\tilde{t}_{\fin}>0$ unless $\Delta\sigma_{x}=0$:
consistently, the connection time cannot
vanish except for the ``configurationally-static'' case. The optimal
temperature evolution follows from
Eq.~\eqref{eq:second-principle-particular-case},
\begin{equation}\label{eq:theta-evol-adiab}
  \tilde{\theta}(\tau)=
  \dfrac{1+\tau\, \Delta\left(\theta \sigma_{x}^{2}\right)}
    {1+\tau\, \Delta\left(\sigma_{x}^{2}\right)}
  ,
\end{equation}
Eqs.~\ref{eq:sigma2-evol-adiab} and \ref{eq:theta-evol-adiab} are
valid in the whole time interval $t\in[0,\tilde{t}_{\fin}]$ or
$0\leq\tau\leq 1$. Both $\tilde{\sigma}_{x}$ and $\tilde{\theta}$ are
monotonic functions of time, the sign of their derivatives being those
of $\Delta \sigma_{x}$ and $\Delta\theta$, respectively. The optimal
potential $\tilde{u}(x,\tau)$ is obtained after inserting
$\tilde{\sigma}_{x}$ and $\tilde{\theta}$ into
Eq.~\eqref{eq:potential-adiab-job}. This expression holds only for
$t\in(0,\tilde{t}_{\fin})$ because $\dot{\tilde{\sigma}}_{x}\neq 0$
for $t=0,t_{\fin}$~\footnote{As found in other problems, the optimal
  control has finite jumps at $t=0$ and $t=t_{\fin}$
  \cite{band_finite_1982,schmiedl_optimal_2007,aurell_optimal_2011,solon_phase_2018,plata_optimal_2019}.
  Adiabaticity is not broken: there is no \textit{instantaneous} heat
  transfer at $t=0$ and/or $t=t_{\fin}$.}


We complement the study above with the analysis of the harmonic case
having time-dependent stiffness
$u(x,\tau)=\frac{1}{2}\kappa(\tau)x^{2}$, a standard experimental
situation. Therein, $p(x,\tau)$ remains Gaussian for all
times, which guarantees shape preservation as shown in Appendix \ref{sec:evol}. With our choice of units, $\kappa_{\ini}=1$ and $\kappa_{\fin}=\theta_{\fin}/\sigma_{x,\fin}^{2}$.
The adiabatic inequality is $\theta_{\fin}^{2}/\kappa_{\fin}\geq 1$.
Eq.~\eqref{eq:potential-adiab-job} gives the relation 
between $\kappa(\tau)$ and $\sigma_{x}(\tau)$, which reduces to~\footnote{This
  expression ensures adiabaticity $d\overline Q=0$, which simplifies
  to $d\theta+\kappa d\sigma_{x}^{2}=0$ for harmonic confinement as said in Appendix \ref{sec:non-opt}.
}
\begin{equation}\label{eq:kappa-harmonic}
  \kappa(\tau)=-\frac{1}{t_{\fin}}\frac{d\ln\sigma_{x}(\tau)}{d\tau}+
  \frac{\theta(\tau)}{\sigma_{x}^{2}(\tau)}. 
\end{equation} 
After some simple algebra, the optimal stiffness results
\begin{equation}\label{eq:kappa-evol-adiab}
  \tilde{\kappa}(\tau)=\frac{C}{\tilde{\sigma}_{x}^{4}(\tau)}, \quad C=\frac{\Delta\theta}{\Delta(\sigma_{x}^{-2})}=
\frac{\Delta\theta}{\Delta(\kappa/\theta)}.
\end{equation}
\begin{figure*}
  \centering
  \includegraphics[height=6cm]{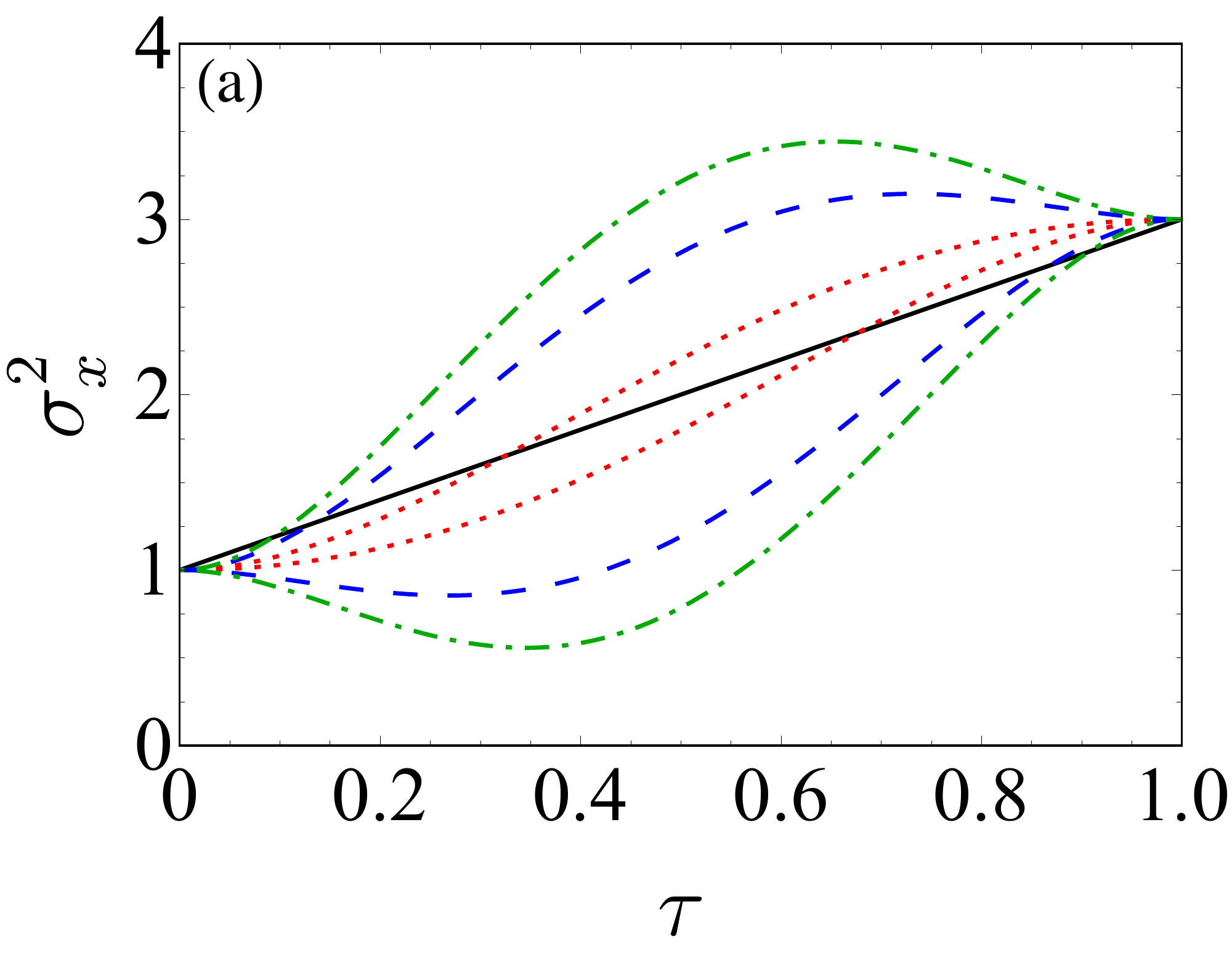} 
  \includegraphics[height=6cm]{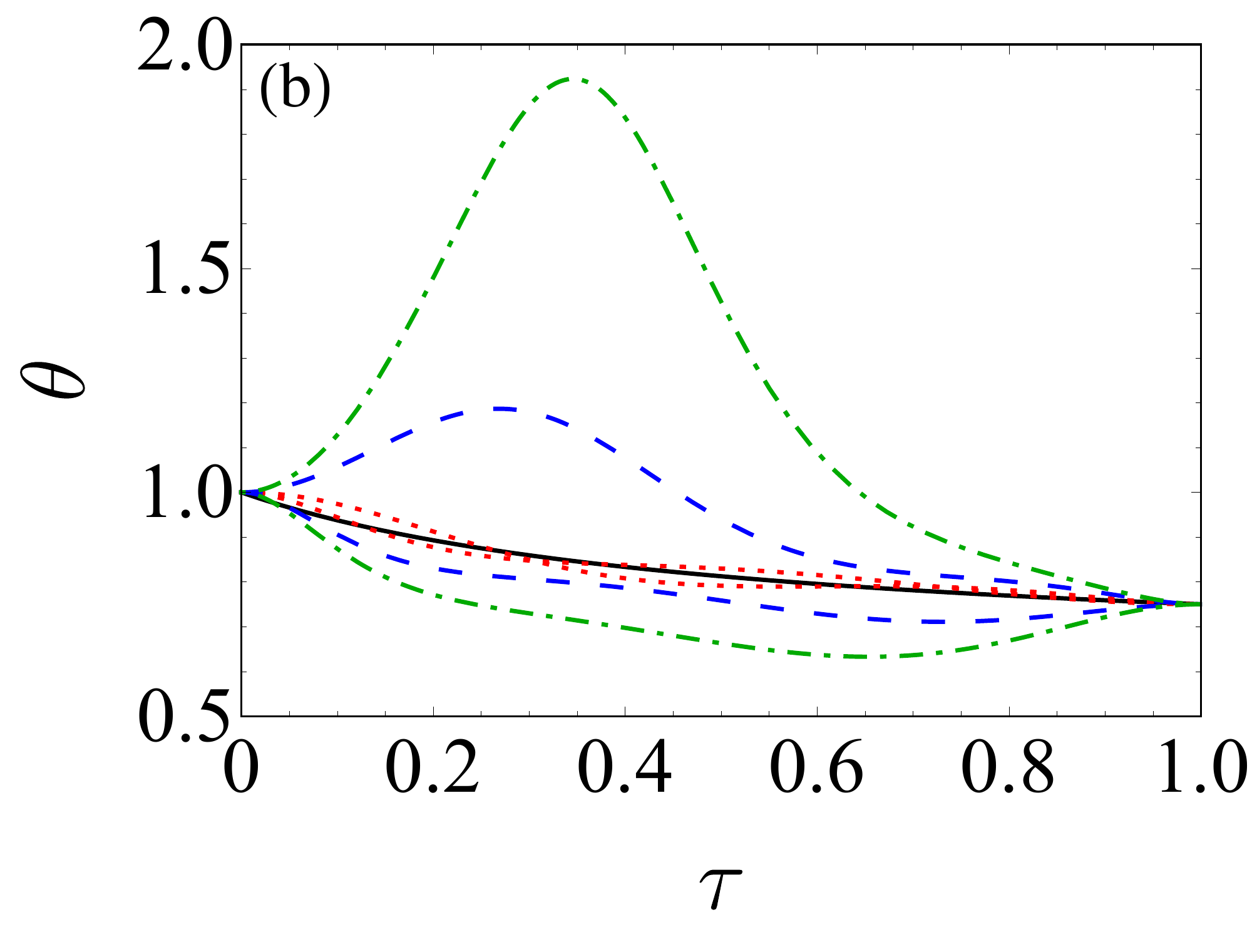}\\
  \includegraphics[height=6cm]{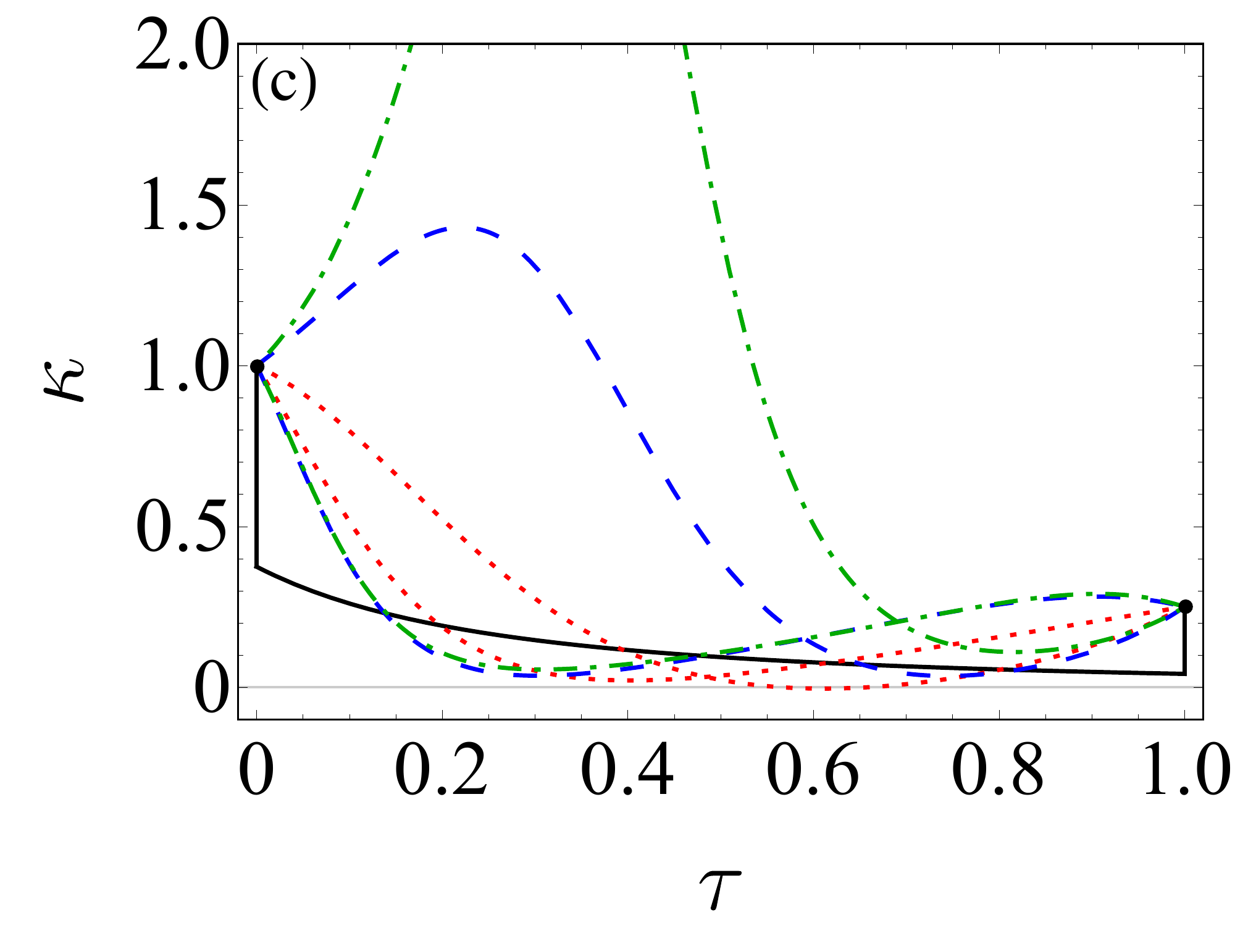}
  \includegraphics[height=6cm]{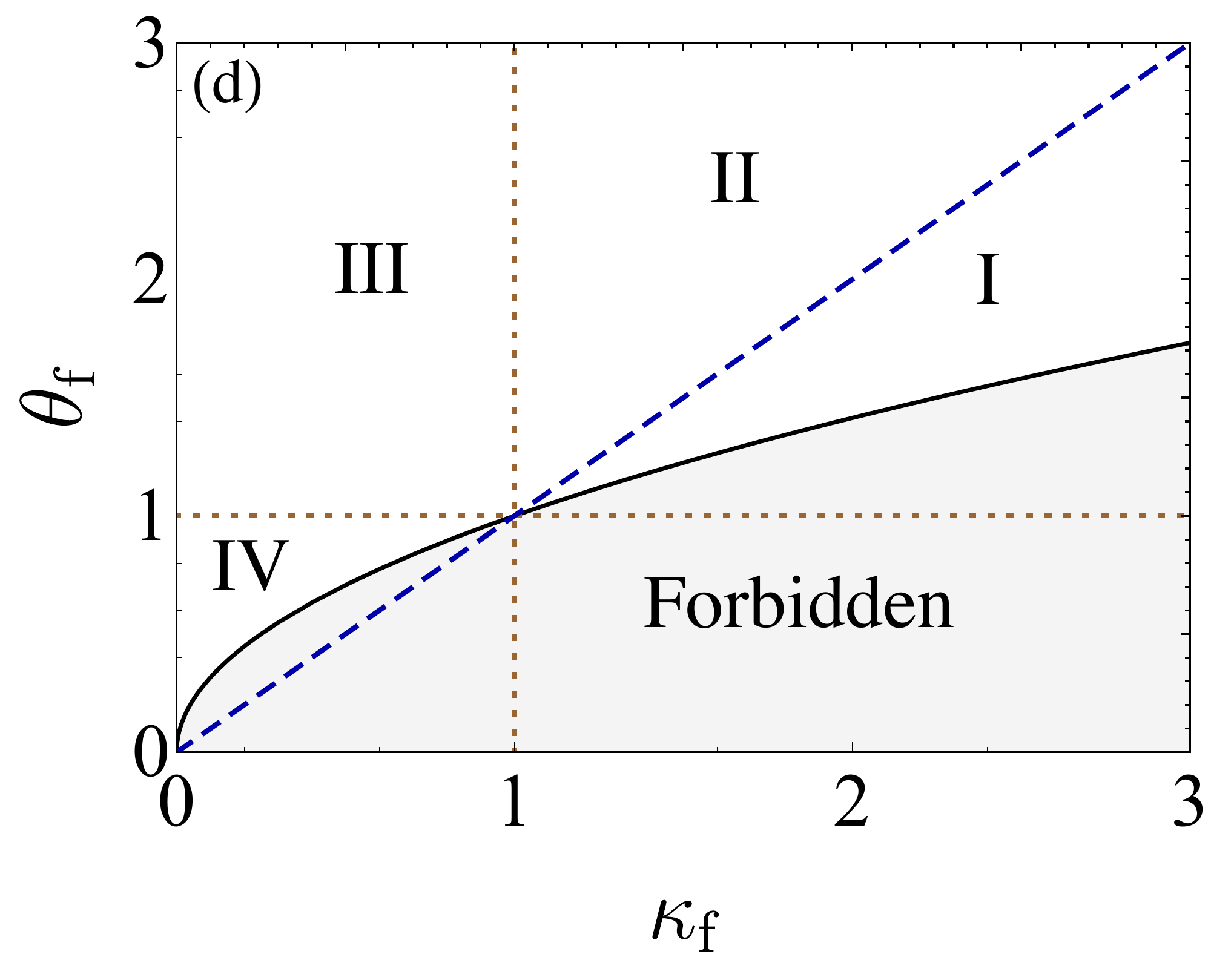}
\caption{\label{fig:profiles} Time evolution of
  $(\sigma_{x}^{2},\theta,\kappa)$ in an adiabatic process (panels
  (a)-(c)) and phase diagram in the $(\kappa_{\fin},\theta_{\fin})$
  plane (panel (d)). The target point on panels (a)-(c) is
  $(\sigma_{x,\fin}^{2}=3,\theta_{\fin}=0.75,\kappa_{\fin}=0.25)$.
  Optimal evolutions (solid lines) give the shortest connecting time
  $\tilde{t}_{\fin}=1.6$. Non-optimal evolutions correspond to longer
  connecting times $t_{\fin}=r\tilde{t}_{\fin}$, with $r=1.25$ (dotted
  red), $r=2$ (dashed blue) and $r=3$ (dot-dashed green).  In (d), the
  solid curve
  $\theta_{\fin}=\theta_{\fin}^{\infty}=\sqrt{\kappa_{\fin}}$
  demarcates the greyed region, which cannot be reached with an
  adiabatic process (``Forbidden''). Reachable points lie on four
  regions, labelled from I to IV. The diagonal line
  $\theta_{\fin}=\kappa_{\fin}$ separates compression
  ($\sigma_{x,\fin}<1$, I) and expansion ($\sigma_{x,\fin}>1$, II-IV)
  regions. The horizontal line separates heating ($\theta_{\fin}>1$,
  I-III) and cooling ($\theta_{\fin}<1$, IV), whereas the vertical one
  separates stiffening ($\kappa_{\fin}>1$, I-II) from loosening
  ($\kappa_{\fin}<1$, III-IV).}
\end{figure*}

Time evolutions of the state point
$(\sigma_{x}^{2}(\tau),\theta(\tau),\kappa(\tau))$, in both optimal
(solid lines) and non-optimal adiabatic processes, are illustrated on
panels (a)-(c) of Figure~\ref{fig:profiles}.  Optimal evolutions are
obtained by particularising Eqs.~\ref{eq:sigma2-evol-adiab},
\ref{eq:theta-evol-adiab} and \ref{eq:kappa-evol-adiab} for each case. Non-optimal evolutions are obtained starting from a
fourth-order polynomial for the variance
$\sigma_{x}^{2}(\tau)=1+b\tau+c\tau^{2}+d\tau^{3}+e\tau^{4}$,
similarly to the approach in Ref.~\cite{martinez_engineered_2016} for
isothermal processes. The values $(b,c,d,e)$ are chosen to fulfil the
boundary conditions for $(\sigma_{x}^{2},\theta,\kappa)$ and the
desired connecting time $t_{\fin}=r\tilde{t}_{\fin}$.  For each $r$
value, there are two paths that connect the initial and final states, see Appendix \ref{sec:non-opt}.  

Figure~\ref{fig:profiles}(d) shows a phase diagram in the plane of
final states $(\kappa_{\fin},\theta_{\fin})$--recall that
$\sigma_{x,\fin}^{2}=\theta_{\fin}/\kappa_{\fin}$.  Over the
  reversible line
  $\theta_{\fin}=\theta_{\fin}^{\qs}=\sqrt{\kappa_{\fin}}$, the
  denominator $\tilde t_\fin$ in Eq.~\eqref{eq:sigma2-evol-adiab} vanishes
  and the minimum time $\tilde{t_{\fin}}$ diverges. The bath
  always must be heated to get compression (region I), whereas the
  trap must be loosened to allow for cooling (IV). However, at odds
  with the isothermal case, the signs of $\Delta\kappa$ and
  $\Delta\sigma_{x}^{2}$ are not univocally related in an adiabatic
  process: stiffening the trap may lead to expansion (II).  Loosening
  entails expansion but the bath may need to be heated (III).

\begin{figure}
  \centering
  \includegraphics[height=6cm]{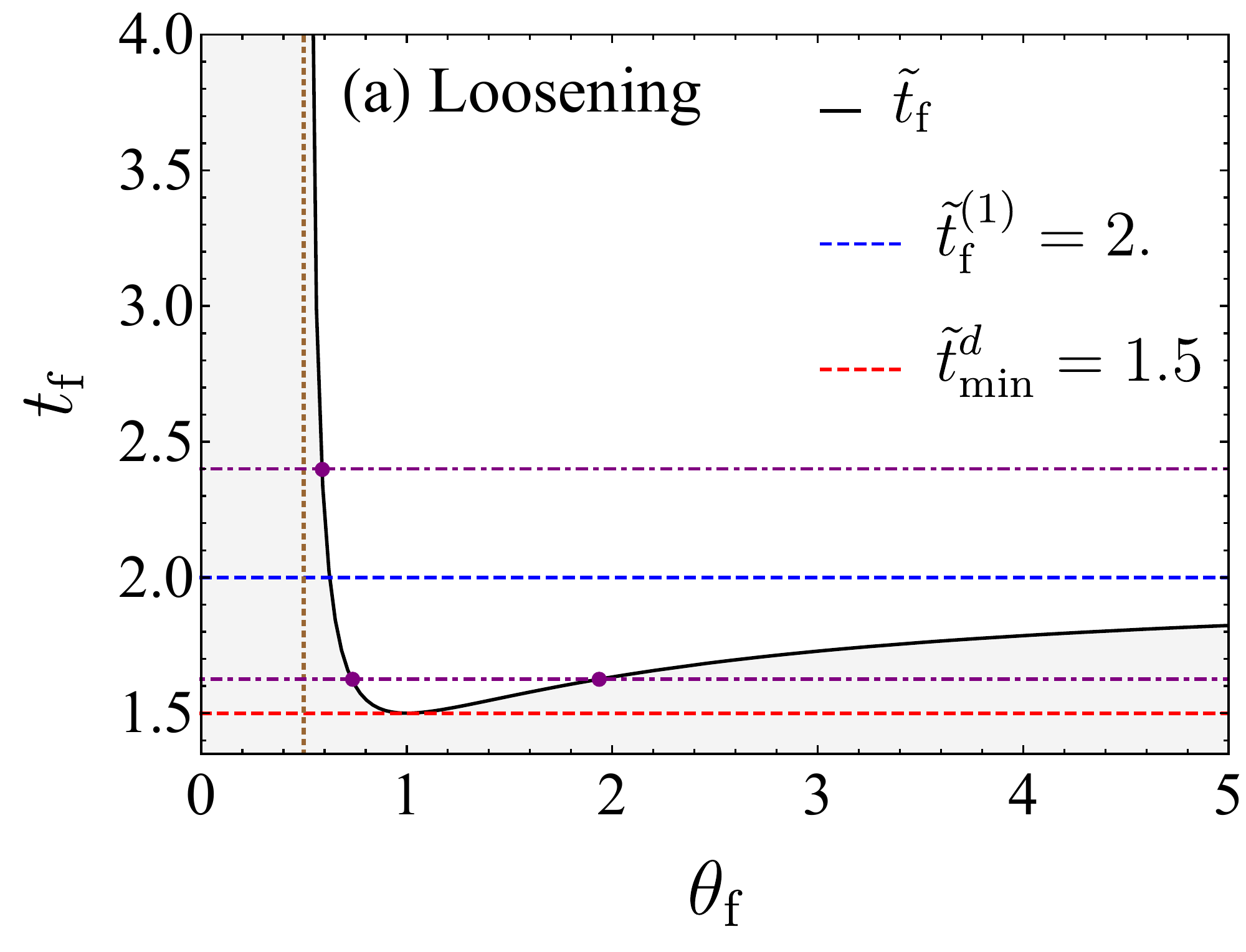}
  \includegraphics[height=6cm]{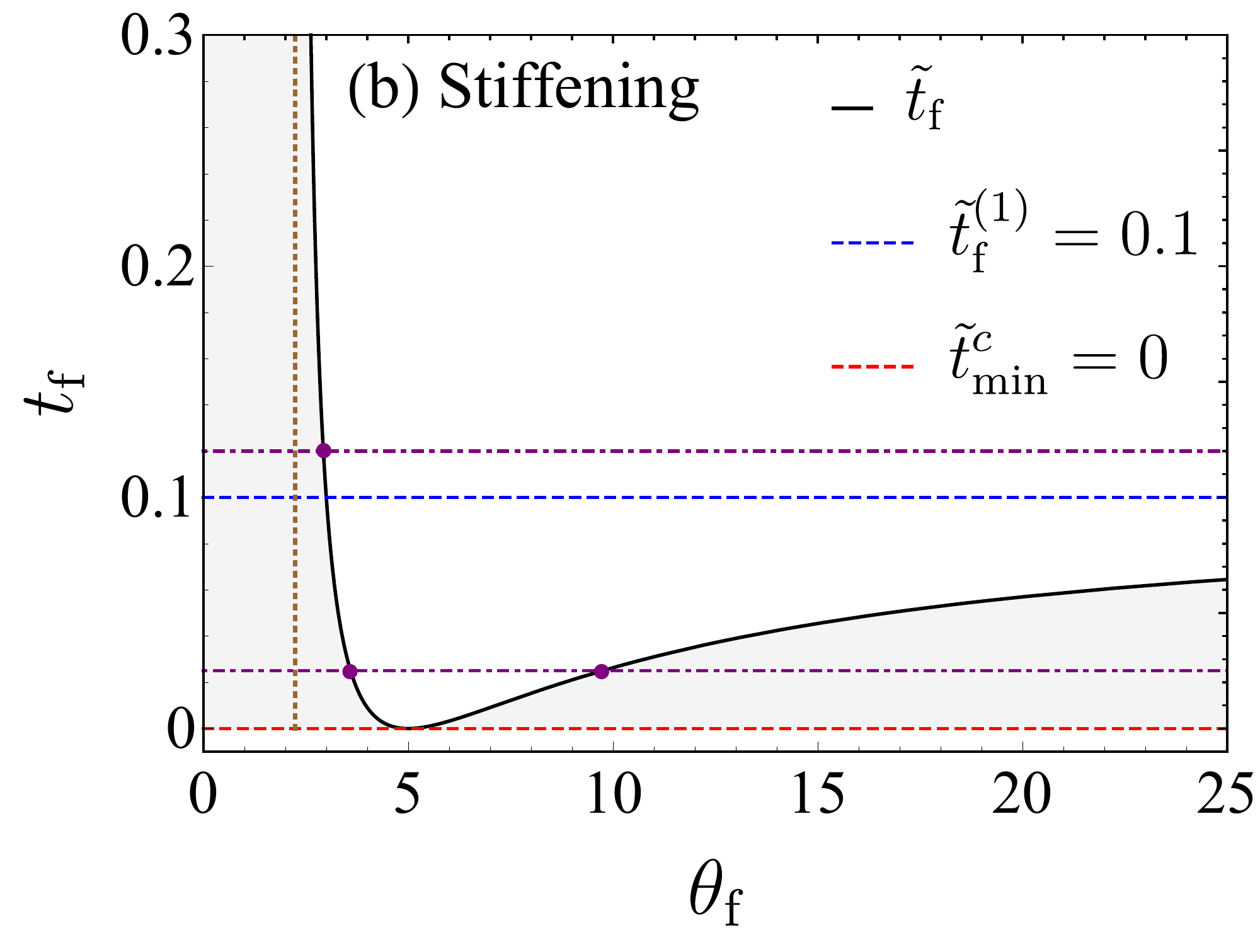}
  \caption{\label{fig:tf-f-of-theta} Minimum connecting time as a
    function of the target temperature, as given by
    Eq. \ref{eq:sigma2-evol-adiab}. Two values of the target stiffness
    are considered: (a) $\kappa_{\fin}=0.25$ and (b)
    $\kappa_{\fin}=5$. The greyed area corresponds to the forbidden
    region $t_{\fin}<\tilde{t}_{\fin}$. On both panels,
    $\tilde{t}_{\fin}$ is non-monotonic and displays an absolute
    minimum $\tilde{t}_{\min}^{\dorc}$ at temperatures
    $\theta_{\fin}^{d}=1$ and $\theta_{\fin}^{c}=\kappa_{\fin}$.  Note
    that $\tilde{t}_{\min}^{d}\neq 0$ whereas
    $\tilde{t}_{\min}^{c}=0$: it is impossible to engineer an
    instantaneous adiabatic process when loosening.  }
\end{figure}

We turn to the characterisation of the minimum time.  For both
  loosening and stiffening, $\tilde{t}_{\fin}$ is a non-monotonic
  function of $\theta_{\fin}$ for fixed $\kappa_{\fin}$;
  $\tilde{t}_{\fin}$ decreases from infinity for the quasi-static
  value $\theta_{\fin}=\theta_{\fin}^{\qs}$ to its minimum
  $\tilde{t}_{\min}^{\dorc}$ at $\theta_{\fin}=\theta_{\fin}^{\dorc}$
  and increases therefrom to $t_{\fin}^{(1)}=(2\kappa_{\fin})^{-1}$ at
  large $\theta_{\fin}$, see Fig.~\ref{fig:tf-f-of-theta}. For
  loosening, $\theta_{\fin}^{d}=1$ ($\Delta\theta=0$) and
  $\tilde{t}_{\min}^{d}=(2\kappa_{\fin})^{-1}-1/2>0$. For stiffening,
  $\theta_{\fin}^{c}=\kappa_{\fin}$ and $\tilde{t}_{\min}^{c}=0$.  The
  horizontal dashed red line marks the minimum time
  $\tilde{t}_{\min}^{\dorc}$, the horizontal blue dashed line the
  asymptotic value $t_{\fin}^{(1)}$, and the dotted vertical asymptote
  the minimum temperature
  $\theta_{\fin}^{\qs}$.

Instead of fixing the final temperature $\theta_{\fin}$, we can fix
the connecting time $t_{\fin}$ and investigate the range of reachable
final temperatures. For instance, a question of experimental relevance
for stiffening is: what is the minimum final fluid temperature for a
given $t_\fin$?  Interestingly, Fig.~\ref{fig:tf-f-of-theta} yields
the answer, if read ``horizontally'' rather than ``vertically'' as
before.  A fresh look at either panel shows that for ``long''
connecting times $t_{\fin}\geq t_{\fin}^{(1)}$, temperatures below the
only one verifying
$\tilde{t}_{\fin}(\tilde{\theta}_{\fin},\kappa_{\fin})=t_{\fin}$ are
inaccessible, because they demand a longer $t_{\fin}$.  This is
illustrated with the horizontal dot-dashed line above
$t_{\fin}^{(1)}$, where $\tilde{\theta}_{\fin}$ is marked with a
purple circle.  For ``short'' connecting times,
$\tilde{t}_{\min}^{d\;{\rm or}\;c}\leq t_{\fin}< t_{\fin}^{(1)}$,
there are two temperatures verifying
$\tilde{t}_{\fin}(\tilde{\theta}_{\fin},\kappa_{\fin})=t_{\fin}$, as
exemplified by the horizontal dot-dashed line below $t_{\fin}^{(1)}$:
only the temperatures between the two purple circles can be
reached. Both the minimum time for fixed final state and the extremal
temperature(s) for fixed connection time can be obtained by means of a
variational approach, as shown in Appendix \ref{sec:optimisation-both}.

For a quasi-static--not necessarily adiabatic--process, the PDF of the
work is delta-peaked around its average value. The heat distribution
is more complex and has been explicitly obtained for the harmonic
potential, being asymmetric around its
mean~\cite{martinez_adiabatic_2015}. For finite-time operation,
calculating these PDFs is far more challenging because position values
at different times are correlated.  Yet, the dominant (up to order of
$1/t_{\fin}$) contributions to the variance of work and heat can be
obtained for slow driving. Work is Gaussian-distributed with variance
\begin{equation}\label{eq:W.fluct}
  \sigma_{W}^{2}\sim\frac{1}{2t_{\fin}}\int_{0}^{1}d\tau\,
  \dot{\kappa}^{2}(\tau)\frac{\theta^2(\tau)}{\kappa^3(\tau)}.
\end{equation}
The change in the heat PDF is also more complex: shape is not conserved and the variance shift is
\begin{equation}
\sigma_{Q}^{2}-(\sigma_{Q}^{\qs})^{2}\sim\sigma_{W}^{2}-\frac{1}{2t_{\fin}}
\Delta\left(\dot{\kappa}\frac{\theta^2}{\kappa^2}\right).
\end{equation}
The last term in $\sigma_{Q}^{2}$ stems from the cross-correlation
between work and heat. For a detailed derivation of work and heat
fluctuations, see Appendix \ref{sec:PDFs-slow}.

\section{Conclusion}
\label{sec:conclusion}

The reported results are very general, being applicable to an
overdamped Brownian particle bound by an arbitrary non-linear
potential. We have shown how two equilibrium states can be connected
with an adiabatic--zero heat--process in a finite time, by explicitly
building such a transformation. The second principle entails the
existence of (i) a forbidden region, i.e. the impossibility of
reaching certain final states from a given initial one, and (ii) a
speed limit for the adiabatic connection, when it is indeed possible:
in general, an instantaneous adiabatic process does not exist.

For compression/decompression of the Brownian particle, further characterisation of these adiabatic transformations can be done. Both the physical discussion and the conclusions stemming from Figs.~\ref{fig:profiles} and \ref{fig:tf-f-of-theta} remain valid for any non-linear potential, by defining a final ``effective'' stiffness $\kappa_{\fin}=\theta_{\fin}/\sigma_{x,\fin}^{2}$ in the non-linear case. Specifically, the emergence of a speed limit--for fixed final state--or a range of reachable temperatures--for fixed connecting time--in adiabatic transformations are robust features of our theory. Also, the phase diagram in Fig.~\ref{fig:profiles}(d) applies to the general non-linear case~\footnote{Even the time dependent profiles of the variance and the temperature, panels (a) and (b) of Fig.~\ref{fig:profiles}, are still valid in the non-linear case. The only result that is specific to the harmonic potential is the time evolution of the stiffness, as given by Eq.~\ref{eq:kappa-harmonic}, in panel (c) of Fig.~\ref{fig:profiles}.}.

In the underdamped case, building finite-time adiabatic processes
remains an open problem. We may surmise though that they cannot be instantaneous,
which would point to the robustness of finite speed limits. An
instantaneous process requires again that the initial and target  PDFs
be coincident, but now for the joint position-velocity PDF. Thus
$\theta_{\fin}=\theta_{\ini}$ and $u_{\fin}(x)=u_{\ini}(x)$: there
would be no room for the entropy to increase, a scenario we can dismiss.

Perspectives concern the stability of the optimal solutions found here
with respect to small perturbations in the trap stiffness, bath
temperature or other constraints.  Besides, our work paves the way for
a theory of control in statistical physics, based on stochastic
thermodynamics. Such optimal solutions clarify the role of
fluctuations and identify fundamental bottleneck with, for instance,
the existence of a speed limit.  The extension of such ideas to
quantum thermodynamics is a promising perspective
\cite{caldeira_path_1983,weiss_quantum_2008,rivas_open_2011}.

\appendix

\section{The overdamped formalism: why?}\label{sec:overdamped}

We are interested in the dynamics of a mesoscopic object in a
suspending fluid (solvent), driven by a time-dependent force field
stemming from the potential $U(X,t)$. The fluid is at equilibrium at
temperature $T$.  For simplicity, we investigate a one dimensional
situation, without loss of generality. The position $X$ of a
`particle' of mass $m$ (a colloid such as a macromolecule, or, at a
much smaller scale, a large molecule) obeys the Langevin equation
\begin{equation}
    m \frac{d^2X}{dt^2} \, = \, - \lambda \, \frac{dX}{dt} - U'(X,t) +  \lambda \sqrt{2 D} \, \xi(t)
\label{eq:Langevin_full}
\end{equation}
where $D=kT/\lambda$ is the diffusion coefficient,
and $\xi(t)$ stands for Gaussian white noise of zero
mean and unit variance.  The drag coefficient $\lambda$ originates from viscous friction and reads
\begin{equation}
    \lambda = 6 \pi \eta r ,
\end{equation}
where $\eta$ is the fluid dynamic viscosity and $r$ the particle radius.
The associated time scale $m/\lambda$ governs the equilibration 
of velocity degrees of freedom; it depends on particle size: explicitly through $r$,
and also through $m\propto r^3$. For a micron-size particle in water at room temperature, we find $m/\lambda$ in the range of $10^{-7}\,$s. This largely exceeds the microscopic solvent correlation time, which justifies the Langevin description
with white noise.
A third important scale in the problem is the time $t_r$ needed to diffuse over 
a particle diameter (hence, $r^2 \sim D \, t_r$). It sets the scale of position evolution; on the other hand, $m/\lambda$ is the time scale ruling
velocity relaxation.
While $m/\lambda \propto r^2$, $t_r \propto r^3$, and we have 
$t_r \gg m/\lambda$. For instance, $t_r \simeq 1\,$s for the above
micron-size colloid. The scale separation $t_r \gg m/\lambda$ holds down to to small 
dimensions, being still true for $r$ in the nanometer range. 
This gap makes it possible to simplify Eq. \eqref{eq:Langevin_full}:
as far as positional degrees of freedom are concerned, inertial terms are irrelevant
and we have
\begin{equation}
 \lambda \, \frac{dX}{dt} = - \partial_X U(X,t) +  \lambda \sqrt{2 D} \, \xi(t) .
\label{eq:Langevin_over}
\end{equation}
This yields the overdamped framework, of much relevance for practical applications,
and the starting point of our treatment. 
Protocols that would require consideration of the inertial term in 
\eqref{eq:Langevin_full} would need to involve time scales below a tenth of microsecond for micron-size particles.

\section{Evolution equation for the variance of the position}
\label{sec:evol}

\subsection{From Langevin to Fokker-Planck}

We now address external driving through a 
harmonic potential with stiffness $k$. Both the temperature and the stiffness,
which are externally controlled, may be time-dependent. The 
 Langevin equation \eqref{eq:Langevin_over} for the particle position reads
\begin{equation}
\label{eq:Langevin}
\lambda\frac{dX(t)}{dt} \,= \, - k \, X(t) + \sqrt{2\lambda k_{B}T} \xi(t).
\end{equation}  

The dynamics of the system can be studied using the probability
density function $P(X,t)$ {for finding the Brownian particle at 
position $X$ at time $t$}. Its time evolution is governed by the Fokker-Planck equation
\begin{equation}
\label{eq:FP}
\lambda\, \partial_t P(X,t) = k\, \partial_X\left[ X P(X,t) \right] + k_{B}T \,\partial_X^2 P(X,t).
\end{equation}  
The Langevin equation
\eqref{eq:Langevin} and the Fokker-Planck equation \eqref{eq:FP} are
equivalent; both completely characterise the time evolution
of the Brownian particle position--mathematically, the stochastic process~\cite{van_kampen_stochastic_1992}.

In light of the above, we can obtain the time evolution of all the
moments or, alternatively, all the cumulants of the position from
either Eq.~\eqref{eq:Langevin} or Eq.~\eqref{eq:FP}. If the initial
condition $P(X,0)$ is Gaussian, as is the case if the system starts
from the corresponding equilibrium state, $P(X,t)$ remains Gaussian
for all times: the two first cumulants, i.e. position's average
$\langle X\rangle$ and variance
$\sigma_{X}^{2}=\langle X^{2}\rangle-\langle X\rangle^{2}$, completely
characterise the evolution of the Brownian particle. 
This can be readily understood from the
Fokker-Planck equation by going to Fourier space. This is the route we
take in the following.

\subsection{Evolution of moments}

First, we define the Fourier transform of $P(X,t)$ as
\begin{equation}
  \label{eq:G(s)}
  G(s,t)\equiv\langle e^{isX}\rangle= \int_{-\infty}^{+\infty}dX\, e^{isX}P(X,t).  
\end{equation}
Therefore, taking the Fourier transform in Eq.~\eqref{eq:FP} leads to
\begin{equation}
\label{eq:FP-G}
\lambda\, \partial_t G(s,t) = - k\, s\, \partial_s G(s,t) - k_{B}T s^2 G(s,t).
\end{equation}  
On the one hand, the expansion of $G(s,t)$ generates the moments
$\mu_{n}(t) \equiv \langle X^n \rangle (t)$, since
$G(s,t)=\sum_{n=0}^\infty (is)^n \mu_n(t)/n!$.  On the other hand, the
expansion of $\ln G(s,t)$ generates the cumulants $\chi_n(t)$,
\begin{equation}\label{eq:lnG}
\ln G(s,t)=\sum _{n=1}^\infty \frac{(is)^n}{n!} \chi_n(t).
\end{equation}  
We have $\chi_1=\mu_1$ (the mean) and $\chi_2=\mu_2-\mu_1^2$ (the variance).

Equation~\eqref{eq:FP-G} can be rewritten as 
\begin{equation}
\label{eq:FP-lnG}
\lambda\, \partial_t \ln G(s,t) = - k\, s\, \partial_s\ln G(s,t) - k_{B}T s^2.
\end{equation}
Introducing the expansion \eqref{eq:lnG} into \eqref{eq:FP-lnG} and
equating the coefficients sharing the same power of $s$, the equations
for the cumulants are obtained as
\begin{equation}
\label{eq:cum-ev}
\lambda \,\frac{d\chi_n(t)}{dt} = -n k\, \chi_n(t) + 2k_{B}T \,\delta_{n,2}, \quad
n\geq 1.
\end{equation} 
{First, we consider the equation for $n=1$. Its solution is
\begin{equation}
  \mu_{1}(t)\equiv\langle X\rangle(t)= \langle
  X\rangle(0)
  \,\exp\!\left[-\frac{1}{\lambda}\int_{0}^{t} dt' k(t')\right].
\end{equation}
Then, the average position remains zero for all times if it is so
initially. Second, the equation for $n=2$ gives the time evolution of
the variance $\chi_{2}\equiv\sigma_{X}^{2}$,
\begin{equation}\label{eq:chi2}
  \lambda\,\frac{d}{dt}\sigma_{X}^{2}=-2k\,\sigma_{X}^{2}+2k_{B}T .
\end{equation}
Third, the equations for $n\geq 2$ entail that an initially Gaussian
distribution remains Gaussian for all times: if $\chi_{n}(0)=0$ for all
$n\geq 2$, we have that $\chi_{n}(t)=0$ for all $n\geq
2$. Eq.~\eqref{eq:chi2} can be solved, with the result
\begin{widetext}
\begin{align}
  \sigma_{X}^{2}(t)=&\sigma_{X,\eq}^{2}(t)+
  \left[\sigma_{X}^{2}(0)-\sigma_{X,\eq}^{2}(0)\right]
 \exp\!\left[-\frac{2}{\lambda}\int_{0}^{t} dt' k(t)'\right]
                      -\int_{0}^{t}dt'\,\frac{d\sigma_{X,\eq}^{2}(t')}{dt'}
                      \exp\!\left[-\frac{2}{\lambda}\int_{t'}^{t} dt'' k(t'')\right], \label{eq:chi2-sol} \\
  \sigma_{X,\eq}^{2}(t)\equiv& \frac{k_{B}T(t)}{k(t)}.
\end{align}
\end{widetext}
If the stiffness of the trap $k$ and the temperature of the fluid $T$
are time-independent, the third term on the rhs is not present and
$\sigma_{X}^{2}(t)$ decays exponentially towards its equilibrium value
$\sigma_{X,\eq}^{2}$.

For the discussion that follows, we
introduce dimensionless variables as in the main text,
\begin{equation}\label{eq:non-dim}
\kappa=\frac{k}{k_{\ini}}, \quad \theta=\frac{T}{T_{\ini}}, \quad
x=\frac{X}{(\sigma_{X,\eq})_{\ini}},
\end{equation}
except for dimensionless time that is defined as
\begin{equation}
  t^{*}=k_{\ini}t/\lambda.
\end{equation}
Note that, consistently with our notation in the paper,
$t_{\fin}^{*}$ is the connection time in the $t^{*}$
variable. Therefore, $\tau=t^{*}/t_{\fin}^{*}$ is the dimensionless
time scale that we have employed throughout the main text. In these
appendices, we will make use of both time scales, $t^{*}$
and $\tau$, depending on which is most useful for each  
situation. In agreement with the notation followed
in the paper, we drop the asterisk for simplicity. Also, for the sake of consistency, $\dot{}\equiv d/d\tau$ and
thus we explicitly write $d/dt$ for derivatives in the time scale $t$.

In dimensionless variables, the evolution equation  of the variance is
given by
\begin{equation}
  \label{eq:variance-eq-dimless}
  \frac{d\sigma_{x}^{2}}{dt}=-2\kappa(t)\sigma_{x}^{2}+2\theta(t).
\end{equation}
The equilibrium variance of the position is
\begin{equation}
  \sigma_{x,\eq}^{2}(t)= \frac{\theta(t)}{\kappa(t)},
\end{equation}
and the time evolution of the mean and variance of the position are
\begin{align}
  \langle x\rangle(t)=&\langle
  x\rangle(0)\,e^{-\varphi(t,0)},
  \label{eq:average-sol-nondim}\\
  \sigma_{x}^{2}(t)=&\sigma_{x,\eq}^{2}(t)+
  \left[\sigma_{x}^{2}(0)-\sigma_{x,\eq}^{2}(0)\right]
                      e^{-2\varphi(t,0)} \nonumber \\
  &-\int_{0}^{t} dt' \,
                         \frac{d\sigma_{x,\eq}^{2}(t')}{dt'} 
                      e^{-2\varphi(t,t')}, \label{eq:chi2-sol-nondim}
\end{align}
where we have defined
\begin{equation}\label{eq:phi-def}
  \varphi(t_{2},t_{1})=\int_{t_{1}}^{t_{2}} dt\, \kappa(t).
\end{equation}
}

\section{Non-optimal adiabatic processes with fourth-order polynomial in the variance} \label{sec:non-opt}

Herein, we describe the non-optimal adiabatic protocols
considered in Fig.~\ref{fig:profiles}(a)-(c) of the paper. {For a
general compression/decompression, the
adiabaticity condition $\dot{\overline Q}=0$ reduces to
$\sigma_{x}^{2}d\theta+\langle xu'\rangle d\sigma_{x}^{2}=0$, where
$u$ is the binding potential. For the harmonic case, $u'=\kappa x$ and
$\langle xu'\rangle=\kappa\sigma_{x}^{2}$, so that adiabaticity
is further simplified to $d\theta+\kappa d\sigma_{x}^{2}=0$
or $\kappa=-\dot{\theta}/(2\sigma_{x}\dot{\sigma}_{x})$, which
provides us with the stiffness.}

The construction of these protocols follows the recipe described in the main
text: starting from a given time-dependence for the variance, first we
compute the time duration for the process, second the temperature
protocol, and finally the stiffness protocol. Specifically, our starting point here 
is a fourth-order polynomial for the time
evolution of the variance,
\begin{equation}
\sigma_{x}^{2}(\tau)=1+b\tau+c\tau^{2}+d\tau^{3}+e\tau^{4},
\end{equation}
which satisfies the initial condition
$\sigma_{x}^{2}(0)=1$. The set of parameters $(b,c,d,e)$ are chosen by
imposing:
\begin{itemize}
\item Given final value for the variance, $\sigma_{x}^{2}(\tau=1)=\sigma_{x,\fin}^{2}$, which leads to
\begin{equation}
\label{eq:Simp1}
1+b+c+d+e=\sigma_{x,\fin}^{2}.
\end{equation}
This constraint (i) reduces the degrees of freedom of our polynomial
from $4$ to $3$ and (ii) is necessary for the consistency of the
proposed protocol, i.e. that which connects the initial and final states.
\item Fixed value of the connecting time $t_{\fin}$, which we give in
  terms of the minimum time $\tilde{t}_{\fin}$ as
  $t_{\fin}=r\tilde{t}_{\fin}$. For our specific shape of
  $\sigma_{x}^{2}(\tau)$, the functional $J[\sigma_{x}]$ in
  Eq.~\eqref{eq:tf-particular-case} reduces to a function $J(b,c,d,e)$
  of the polynomial parameters. Thus,
  Eqs.~\eqref{eq:tf-particular-case} with the condition
  $t_{\fin}=r\tilde{t}_{\fin}$ entails that
\begin{equation}
\label{eq:Simp2}
J(b,c,d,e)= r (\Delta \sigma_x^2)^2.
\end{equation}
Note that this condition ensures that the temperature protocol
$\theta(\tau)$ obtained from
Eq.~\eqref{eq:second-principle-particular-case} verifies the boundary
conditions for both the initial and final times, $\theta(\tau=0)=1$
and $\theta(\tau=1)=\theta_{\fin}$.
\item Continuity in the stiffness protocol at both the initial and
  final times, i.e. $\kappa(0)=1$ and
  $\kappa(\tau=1)=\kappa_{\fin}$. Following our discussion above,
\begin{equation}
\label{eq:Simp3}
-\left. \frac{\dot{\theta}}{2\dot\sigma_{x}}\right|_{\tau=0}=1, \quad
-\left. \frac{\dot{\theta}}{2\dot\sigma_{x}}\right|_{\tau=1}=
\frac{\theta_{\fin}}{\sigma_{x,\fin}}
.
\end{equation}

\end{itemize}

The system of equations Eqs. \eqref{eq:Simp1}-\eqref{eq:Simp3} can be
exactly solved and provides us with two sets of parameters
\begin{equation}
b=0, \quad 
c=\frac{1}{2}\left(6\Delta \sigma_{x}^{2} \pm \Gamma  \right) ,\quad
d=-2\Delta\sigma_{x}^{2} \mp \Gamma ,\quad 
e= \pm \frac{1}{2} \Gamma 
\end{equation}
where the up and down signs correspond to the first and second
solutions, respectively, and we have introduced
\begin{equation}
\Gamma=\sqrt{42(5r-6)(\Delta \sigma_{x}^{2})^2}.
\end{equation} 
Thus, these non-optimal protocols are limited to $r\geq 6/5$ and allow
us to obtain connection times that are, at least, $20\%$ longer than
the minimum time $\tilde{t}_{\fin}$. This restriction stems from our
imposing of continuous stiffness at the boundaries, as given by
Eq.~\eqref{eq:Simp3}. Had we relaxed this condition, we would 
have obtained a larger set of solutions for the parameters $(b,c,d,e)$
including the optimal solution $(\Delta y,0,0,0)$ for $r=1$, whose
associated optimal stiffness has finite jumps at the boundaries, as
discussed in the main text.

\section{Optimisation problems}\label{sec:optimisation-both}

\subsection{Optimal (extremal) temperature for fixed running time}
\label{sec:optim}

\subsubsection{Statement of the variational problem}

We would like to minimise the final temperature in an adiabatic
process for the trapped Brownian particle. Therefore, {consider} the
temperature increment
\begin{equation}
  \label{eq:delta-theta}
  \Delta\theta\equiv\theta_{\fin}-\theta_{\ini}=\int_{0}^{t_{\fin}}dt\, \frac{d\theta}{dt} .
\end{equation}
This is a ``constrained'' minimisation problem, since we seek the
minimisation of $\Delta\theta$ that is compatible with (i) the time
evolution of the variance of the Brownian particle,
Eq.~\eqref{eq:chi2}, and (ii) the adiabaticity condition,
$d\theta+\kappa\, d\sigma_{x}^{2}=0$, i.e.
\begin{equation}\label{eq:auxiliary}
  \frac{d\sigma_x^2}{dt}=-2\kappa \sigma_x^2+2\theta, \qquad \kappa\frac{d\sigma_x^2}{dt}+\frac{d\theta}{dt}=0.
\end{equation}
Therefore, we have to introduce Lagrange multiplier functions
$\lambda(t)$ and $\mu(t)$ ensuring that the above conditions hold for
all times, as explained in Ref.~\cite{lanczos_variational_1970} for
minimisation problems with ``auxiliary conditions''--or in
Ref.~\cite{gelfand_calculus_2000} for minimisation problems with
``subsidiary conditions''.

Throughout this section, we use the abbreviation
$y\equiv\sigma_{x}^{2}$ to simplify the notation. Then, we
look for functions that make
\begin{align}
  \calS[y,\kappa,\theta,\lambda,\mu]=&\int_{0}^{t_{\fin}}dt\, \frac{d\theta}{dt}+\int_{0}^{t_{\fin}}
  dt\,\lambda(t) \left(\frac{dy}{dt}+2\kappa
                                        y-2\theta\right) \nonumber \\
  & +\int_{0}^{t_{\fin}}
    dt\,\mu(t)\left(\kappa\frac{dy}{dt}+\frac{d\theta}{dt}\right)
    \label{eq:action-1}
\end{align}
stationary. We have to minimise the ``action''
\begin{equation}
  \label{eq:action-2}
  \calS[y,\kappa,\theta,\lambda,\mu]=\int_{0}^{t_{\fin}}dt\, \calL \left(\kappa,y,\frac{dy}{dt},\theta,\frac{d\theta}{dt},\lambda,\mu \right) ,
\end{equation}
in which we have the ``Lagrangian''
\begin{align}
  \label{eq:lagrangian}
  \calL \left( \kappa,y,\frac{dy}{dt},\theta,\frac{d\theta}{dt},\lambda,\mu \right)=&
  \frac{d\theta}{dt}+\lambda \left(\frac{dy}{dt}+2\kappa
                                                                                      y-2\theta\right) \nonumber \\
  &+\mu\left(\kappa\frac{dy}{dt}+\frac{d\theta}{dt}\right).
\end{align}
Note that the ``Lagrangian'' does not depend on $d\kappa/dt$. This
means that the corresponding Euler-Lagrange for $\kappa$ can be used
to eliminate $\kappa$ in favour of the remainder of the
variables~\footnote{The same is true for the Lagrange multipliers
  $\lambda$ and $\mu$, but this is not a peculiarity of the problem
  with which we deal here, but a general property of the Lagrange
  multiplier method: by construction, the Lagrangian never depends on
  the time derivatives of the Lagrangian multipliers.}.

The boundary conditions for the minimisation problem are
the following: (i) given initial equilibrium state, i.e. given values of
  $\kappa_{\ini}$, $y_{\ini}$ and $\theta_{\ini}$.
  \begin{equation}
    \label{eq:bc-t=0}
    {{\kappa(t=0)=\kappa_{\ini}, \; y(t=0)=y_{\ini}, \;
        \theta(t=0)=\theta_{\ini} \, = \, \kappa_{\ini} \, y_{\ini},}}
  \end{equation}
  and (ii) given value of the final stiffness $\kappa_{\fin}$ and
  equilibrium condition at the final time,
  $\kappa_{\fin}y_{\fin}=\theta_{\fin}$.
  \begin{equation}
    \label{eq:bc-t=tf}
    {{\kappa(t=t_{\fin})=\kappa_{\fin}, \quad \kappa(t=t_{\fin})y(t=t_{\fin})=\theta(t=t_{\fin}).}}
  \end{equation}

By taking an infinitesimal variation of $\calS$ and equating it to
zero, not only do we get the Euler-Lagrange equations for the
minimisation problem, but also the adequate boundary conditions--as
discussed in Ref.~\cite{lanczos_variational_1970}, section II.15. The
boundary term in $\delta\calS$ must vanish,
\begin{equation}
  \label{eq:boundary-terms}
  \left.p_{\kappa}\delta\kappa+p_{y}\delta y+p_{\theta}\delta\theta \right|_{0}^{t_{\fin}}=0.
\end{equation}
We have here introduced the canonical momenta, which for
our problem read
\begin{subequations}\label{eq:canon-mom}
  \begin{align}
    \label{eq:pk}
    p_{\kappa}\equiv&\pder{\calL}{(d\kappa/dt)}=0, \\
    \label{eq:py}
    p_{y}\equiv&\pder{\calL}{(dy/dt)}=\lambda+\mu\kappa, \\
    \label{eq:ptheta}
    p_{\theta}\equiv&\pder{\calL}{(d\theta/dt)}=1+\mu.
  \end{align}
\end{subequations}
Since $\kappa_{\ini}$, $y_{\ini}$ and $\theta_{\ini}$ are fixed, there
is no boundary contribution coming from $t=0$. For $t=t_{\fin}$,
{however}, we have a different situation, $\delta\kappa_{\fin}=0$ but
$y_{\fin}$ and $\theta_{\fin}$ are simply linked by the equilibrium
condition, which entails that
$\kappa_{\fin}\delta y_{\fin}=\delta\theta_{\fin}$. Therefore, we have
that
\begin{equation}
  \label{eq:bc-1}
  p_{\kappa\fin}\cancelto{0}{\delta\kappa_{\fin}}+p_{y\fin}\delta
y_{\fin}+p_{\theta\fin}\delta\theta_{\fin}=\left(p_{y\fin}+p_{\theta\fin}
  \kappa_{\fin}\right)\delta y_{\fin}=0,
\end{equation}
so that
\begin{equation}
  \label{eq:bc-2}
  {{p_{y}(t=t_{\fin})+p_{\theta}(t=t_{\fin})
      \kappa(t=t_{\fin})=0,}}
\end{equation}
since $\delta y_{\fin}$ is arbitrary. By employing the expressions for
$p_{y}$ and $p_{\theta}$ found above, we get
\begin{equation}
  \label{eq:bc-3}
  \lambda_{\fin}+\mu_{\fin}\kappa_{\fin}+(1+\mu_{\fin})\kappa_{\fin}=\kappa_{\fin}+\lambda_{\fin}+2\mu_{\fin}\kappa_{\fin}=0
\end{equation}
for the lacking boundary condition, i.e.
\begin{equation}
  \label{eq:bc-4}
  \kappa(t=t_{\fin})+\lambda(t=t_{\fin})+
      2\mu(t=t_{\fin})\kappa(t=t_{\fin})=0.
\end{equation}

\subsubsection{Euler-Lagrange equations}

Now, we write the Euler-Lagrange equations for the minimisation
problem. First, taking into account Eq.~\eqref{eq:pk} and $\partial_{\kappa}\calL=2\lambda y+\mu\,{dy}/{dt}$,
\begin{equation}
  \label{eq:EL-kappa}
  0=2\lambda y+\mu\frac{dy}{dt}.
\end{equation}
Second, we bring to bear Eq.~\eqref{eq:py} and
$\partial_{y}\calL=2\kappa\lambda$,
\begin{equation}
  \label{eq:EL-y}
  \frac{d}{dt}\left(\lambda+\mu\kappa\right)=2\kappa\lambda.
\end{equation}
Third, we make use of Eq.~\eqref{eq:ptheta} and
$\partial_{\theta}\calL=-2\lambda$ to write
\begin{equation}
  \label{eq:EL-theta}
  \frac{d\mu}{dt}=-2\lambda.
\end{equation}
In addition, since by construction the Lagrangian does not depend on
the time derivatives of the Lagrange multipliers ${\lambda}$ and ${\mu}$, the Euler-Lagrange equations for
$\lambda$ and $\mu$ reduce to the constraints or auxiliary
conditions~\eqref{eq:auxiliary}.

It is straightforward to get rid of the Lagrange multipliers by first
inserting Eq.~\eqref{eq:EL-theta} into~\eqref{eq:EL-kappa}, which gives
\begin{equation}
  \label{eq:mu-sol}
  \mu\frac{dy}{dt}-y\frac{d\mu}{dt}=0 \quad \Rightarrow \quad \mu=c_{1}y,
\end{equation}
where $c_{1}$ is an arbitrary constant, to be determined later by
imposing the boundary conditions. Moreover, Eq.~\eqref{eq:EL-theta} yields
\begin{equation}
  \label{eq:lambda-sol}
  \lambda=-\frac{c_{1}}{2}\frac{dy}{dt}.
\end{equation}
These expressions for the multipliers in terms of $y$ and $dy/dt$
allow us to work out the solution, as detailed below. The constant
$c_{1}$ should be non-zero because its vanishing leads to
$\lambda(t)=\mu(t)=0$, i.e. the situation without constraints.

Inserting Eqs.~\eqref{eq:mu-sol} and~\eqref{eq:lambda-sol} into~\eqref{eq:EL-y}, we get
\begin{equation}
  \label{eq:aux-1}
  \frac{d^{2}y}{dt^{2}}-4\kappa\frac{dy}{dt}-2\frac{d\kappa}{dt}y=0,
\end{equation}
after taking into account that $c_{1}\neq 0$. By employing
Eq.~\eqref{eq:auxiliary} to take the time
derivative of the evolution equation for $y$ and make use of the adiabatic condition, it is also shown that
\begin{equation}
  \label{eq:aux-2}
  \frac{d^{2}y}{dt^{2}}+4\kappa\frac{dy}{dt}+2\frac{d\kappa}{dt}y=0.
\end{equation}
Combining Eqs.~\eqref{eq:aux-1} and~\eqref{eq:aux-2}, we obtain 
\begin{equation}
  \label{eq:kappa-y2}
  2\kappa\frac{dy}{dt}+\frac{d\kappa}{dt}y=0 \Rightarrow \quad
  {{\kappa y^{2}=c_{2},}}
\end{equation}
where $c_{2}$ is an arbitrary constant.

Finally, taking into account Eq.~\eqref{eq:kappa-y2}, we find the
expressions for the variance and the temperature. The adiabatic
condition is now simplified to
\begin{equation}
  \label{eq:adiabatic-simplified}
  c_{2}\frac{1}{y^{2}}\frac{dy}{dt}+\frac{d\theta}{dt}=0 \Rightarrow {{\theta=\frac{c_{2}}{y}+\frac{c_{3}}{2},}}
\end{equation}
in which $c_{3}$ is another arbitrary constant--the factor $1/2$ on
the rhs is convenient later. Substituting Eqs.~\eqref{eq:kappa-y2} and~\eqref{eq:adiabatic-simplified} into the evolution equation
for the variance $y$ gives
\begin{equation}
  \label{eq:y-straight-line}
  \frac{dy}{dt}+
  \cancel{2\frac{c_{2}}{y^{2}}y}-2\left(\cancel{\frac{c_{2}}{y}}+
    \frac{c_{3}}{2}\right)=0 \Rightarrow
  {{\frac{dy}{dt}=c_{3},}} \Rightarrow
  {{y=c_{3}t+c_{4}.}}
\end{equation}
Once more, $c_{4}$ is an arbitrary constant.

\subsubsection{Solution of the  problem}

Equations~\eqref{eq:kappa-y2}, \eqref{eq:adiabatic-simplified} and
\eqref{eq:y-straight-line} provide the solution to the minimisation
problem. The constants $(c_{1},c_{2},c_{3},c_{4})$ have to be written in
terms of physical quantities by imposing the boundary conditions. It
may seem odd at first sight that there are $4$ constants but $6$
boundary conditions. The reason is the same as in other problems in
stochastic thermodynamics: $\kappa$ may have jumps at the
boundaries. In the present context, this peculiar behaviour is readily
understood: the conjugate moment $p_{k}=\partial_{\kappa}\calL$
identically vanishes and therefore $\delta\kappa(t=0)$ and
$\delta\kappa(t=t_{\fin})$ are in fact arbitrary when imposing the
extremality condition $\delta\calS=0$. This means that $\kappa$ can
indeed have finite-jump discontinuities at the initial and final
times: $\kappa(t=0^{+})$ and $\kappa(t=t_{\fin}^{-})$ do not coincide in
general with $\kappa_{\ini}$ and $\kappa_{\fin}$.

Following the discussion above, we now impose the four relevant
boundary conditions
\begin{subequations}
  \label{eq:bc-relevant}
\begin{align}
  y(t=0)=&y_{\ini}, \\ \theta(t=0)=&\theta_{\ini}, \\
   \kappa_{\fin}y(t=t_{\fin})=&\theta(t=t_{\fin}), 
  \\ \kappa_{\fin}+\lambda(t=t_{\fin})+2\kappa_{\fin}\mu(t=t_{\fin})=&0.
\end{align}
\end{subequations}
The constants $c_{3}$ and $c_{4}$ are directly obtained as
\begin{equation}
  \label{eq:c3-c4}
  {{c_{3}=\frac{y_{\fin}-y_{\ini}}{t_{\fin}},
      \quad c_{4}=y_{\ini}.}}  
\end{equation}
Note that $y_{\fin}$ does not have a definite value but is related to
$\theta_{\fin}$ by the equilibrium condition; this will be brought to
bear later. The optimal time evolution for the variance is then
\begin{equation}
  \label{eq:y-sol}
  {{y(t)=y_{\ini}+
      \frac{y_{\fin}-y_{\ini}}{t_{\fin}}t.}}
\end{equation}
Now, particularising Eq.~\eqref{eq:adiabatic-simplified} for $t=0$ makes
it possible to obtain $c_{2}$,
\begin{equation}
  \label{eq:c2}
  \theta_{\ini}=\frac{c_{2}}{y_{\ini}}+\frac{c_{3}}{2} \Rightarrow
  \quad {{c_{2}=y_{\ini}\left(\theta_{\ini}-
        \frac{y_{\fin}-y_{\ini}}{2t_{\fin}}\right).}}
\end{equation}
Using again Eq.~\eqref{eq:adiabatic-simplified} but for an arbitrary time
$t$, one gets  after some simple algebra
\begin{equation}
  \label{eq:theta-sol}
 {{\theta(t)=
 \frac{y_{\ini}\theta_{\ini}+\dfrac{(y_{\fin}-y_{\ini})^{2}}{2t_{\fin}^{2}}t}
 {y_{\ini}+\dfrac{y_{\fin}-y_{\ini}}{t_{\fin}}t}.}}
\end{equation}
Substituting $t=t_f$ into this equation, we obtain
\begin{equation}
  \label{eq:adiabatic-inequality-1}
  \theta_{\fin}=\frac{y_{\ini}\theta_{\ini}}{y_{\fin}}+
  \frac{(y_{\fin}-y_{\ini})^{2}}{2t_{\fin}y_{\fin}},
\end{equation}
{so that}
\begin{equation}
  \label{eq:adiabatic-inequality-2}
  \theta_{\fin}\geq \frac{y_{\ini}\theta_{\ini}}{y_{\fin}},
\end{equation}
with the equality holding in the limit as $t_{\fin}\to\infty$.

We have yet to impose the boundary condition
$y_{\fin}=\theta_{\fin}/\kappa_{\fin}$. We do so in
Eq.~\eqref{eq:adiabatic-inequality-1},
\begin{subequations}
  \label{eq:quadratic-eq-1}
  \begin{align}
  \theta_{\fin}=&\frac{\kappa_{\fin}y_{\ini}\theta_{\ini}}{\theta_{\fin}}+
  \frac{\kappa_{\fin}}{2t_{\fin}}\frac{\left(\dfrac{\theta_{\fin}}{\kappa_{\fin}}-y_{\ini}\right)^{2}}{\theta_{\fin}}
\\  \Rightarrow \theta_{\fin}^{2}=&\kappa_{\fin}y_{\ini}\theta_{\ini}+
  \frac{\kappa_{\fin}}{2t_{\fin}}
  \left(\frac{\theta_{\fin}}{\kappa_{\fin}}-y_{\ini}\right)^{2}.
  \end{align}
\end{subequations}
This is a quadratic equation for $\theta_{\fin}$
in terms of the fixed parameters $\kappa_{\fin}$, $y_{\ini}$,
$\theta_{\ini}$, and $t_{\fin}$. Solving it for $t_{\fin}$, we find
\begin{equation}
  \label{eq:solved-for-tf} t_{\fin}=\frac{\left(\dfrac{\theta_{\fin}}{\kappa_{\fin}}-\dfrac{\theta_{\ini}}{\kappa_{\ini}}\right)^{2}}
{2\left(\dfrac{\theta_{\fin}^{2}}{\kappa_{\fin}}-\dfrac{\theta_{\ini}^{2}}{\kappa_{\ini}}\right)},
\end{equation}
which is equivalent to Eq.~\eqref{eq:sigma2-evol-adiab} in the main text,
particularised for the harmonic potential.

It is worth noting that the constant $c_{1}$ has not been necessary to
obtain the solution for the physical quantities, the stiffness
$\kappa(t)$, the variance $y(t)$, and the temperature $\theta(t)$. It is only needed to derive the final expressions for the
Lagrange multipliers $\lambda(t)$ and $\mu(t)$. For the sake of
completeness, we give the expression for $c_{1}$ that follows from
Eq.~\eqref{eq:bc-4},
\begin{equation}
  \label{eq:c1-sol}
  \kappa_{\fin}-\frac{c_{1}c_{3}}{2}+2c_{1}y_{\fin}\kappa_{\fin}=0 \Rightarrow
  \quad c_{1}=\frac{2\kappa_{\fin}}{c_{3}-4\theta_{\fin}}.
\end{equation}

\subsection{Minimum time for fixed initial and final states}
\label{sec:minim}

We turn our attention to another optimisation problem: that of obtaining the
minimum time to connect two given equilibrium states with an adiabatic
process. This problem has been solved in the paper by an ad-hoc
procedure, but it can be addressed in a way similar to the one
employed in the previous section. In this case, we would like to
minimise
\begin{equation}
  \label{eq:tf-def}
  t_{\fin}=\int_{0}^{t_{\fin}}dt\, 1, 
\end{equation}
submitted again to the constraints given by
Eq.~\eqref{eq:auxiliary}. Therefore, we have to minimise a new
``action''
\begin{equation}
  \label{eq:action-3}
  \hat{\calS}[y,\kappa,\theta,\lambda,\mu]=\int_{0}^{t_{\fin}}dt\, \hat{\calL} \left(\kappa,y,\frac{dy}{dt},\frac{d\theta}{dt},\lambda,\mu \right) ,
\end{equation}
in which we have the new ``Lagrangian''
\begin{align}  \hat{\calL}\left(\kappa,y,\frac{dy}{dt},\theta,\frac{d\theta}{dt},
  \lambda,\mu\right)
      =&
         \calL\left(\kappa,y,\frac{dy}{dt},\theta,\frac{d\theta}{dt},\lambda,\mu\right)
         \nonumber \\ &+\frac{d}{dt}\left(t-\theta\right).                                                          \label{eq:lagrangian-2}
\end{align}
Since $\hat{\calL}$ and $\calL$ differ by the total derivative of a
function that depend only on the ``coordinates''--and not on the
velocities, we know that the Euler-Lagrange equations for both
minimisation problems will be the same. Anyhow, we cannot {yet}
conclude that the solution to both problems is the same, since the
boundary conditions for them are not~\footnote{In this particular
  case, the fact that the Euler-Lagrange equations do not change can
  be easily seen, without ``invoking'' the equivalence of Lagrangians
  differing by the time derivative of a function. The equality
  $\hat{\calL}=\calL+1-\dot{\theta}$ implies that all the partial
  derivatives of $\hat{\calL}$ and $\calL$ are equal, except for the
  momentum
  $\hat{p}_{\theta}\equiv\partial\hat{\calL}/\partial\dot{\theta}=p_{\theta}-1$. However,
  since it is the derivative of the momenta that enter into the
  Euler-Lagrange equations, they obviously remain unchanged.}.

In this case, the boundary conditions are simpler than {those addressed in section \ref{sec:optim}}, because $(\kappa,y,\theta)$ have prescribed
values at the initial and final times, although the latter is not
fixed; it is the quantity that we want to
minimise. Specifically, Eq.~\eqref{eq:bc-t=0} and Eq.~\eqref{eq:bc-t=tf}
remain valid but Eq.~\eqref{eq:bc-4} must be substituted with
\begin{equation}\label{eq:bc-for-thetaf}
  \theta(t=t_{\fin})=\theta_{\fin}.
\end{equation}
Therefore, we deal with a ``standard'' variational problem, for which
$\delta\kappa$, $\delta y$ and $\delta\theta$ vanish at the
boundaries, similarly to the situation found in Classical Mechanics.
Notwithstanding, once more we have that $\kappa$ may have finite jump
discontinuities at the boundaries: recall that its corresponding
canonical momentum verifies $\hat{p}_{\kappa}\equiv 0$.

Since the Euler-Lagrange equations are unchanged, Eq.~\eqref{eq:mu-sol},
Eq.~\eqref{eq:lambda-sol}, Eq.~\eqref{eq:kappa-y2},
Eq.~\eqref{eq:adiabatic-simplified} and Eq.~\eqref{eq:y-straight-line} still
hold. In principle, we should have to reobtain the constants
$(c_{2},c_{3},c_{4})$ with the new boundary conditions. However, it is
{readily} realised that the substitution of Eq.~\eqref{eq:bc-4} with
Eq.~\eqref{eq:bc-for-thetaf} leaves their expressions unchanged,
because Eq.~\eqref{eq:bc-4} was not employed in their derivation for the
optimal temperature problem.  The only difference is that
$\theta_{\fin}$ is now fixed and $t_{\fin}$ is the variable being
minimised, instead of the other way round. In light of the previous discussion, it {appears} that the same function
relates the optimal values $\theta_{\fin}$ and $t_{\fin}$ for both
physical situations, as argued in the main text on physical
grounds. 

{\section{Fluctuations of the energy increment, work and
  heat}\label{sec:PDFs-slow}

In the quasi-static limit, the PDFs for the increment of potential
energy $\Delta u$, work $W$ and heat $Q$ have been obtained for the
harmonic
potential~\cite{martinez_adiabatic_2015}.
The calculations rely on the values of $x(t)$ and $x(t')$ being
uncorrelated for all times, which is strictly true only for an
infinite connecting time. Here, we consider how these results are
changed by a finite-time but slow driving, i.e. the situation when the
dimensionless connecting time $t_{\fin}\gg 1$ and both the stiffness
and the temperature are slowly varied, 
i.e. their time derivatives in the
``fast'' scale $t$ is of the order of $t_{\fin}^{-1}$.

\subsection{Time correlations}\label{sec:time-correlations}

For slow drivings, it is convenient to go to the time scale
$\tau=t/t_{\fin}$, over which the evolution equation of the variance of the
position is given by
\begin{equation}\label{eq:variance-eq-tau} t_{\fin}^{-1}\frac{d\sigma_{x}^{2}}{d\tau}=-2\kappa(\tau)\sigma_{x}^{2}(\tau)+
  2\theta(\tau).
\end{equation}
Therefore,
the lowest order approximation for the variance is
\begin{equation}\label{eq:sigmax-lowest}
  \sigma_{x}(\tau)\sim \sigma_{x,\eq}(\tau).
\end{equation}
This expression is uniformly valid in time: it verifies both boundary
conditions in Eq.~\eqref{eq:bc-time} of the main text. Therefore, the
one-time PDF for the position at time $t$ is Gaussian with zero mean
and this variance, i.e.
\begin{equation}\label{eq:one-time-prob}
  p(x,t)\sim \frac{1}{\sqrt{2\pi\sigma_{x,\eq}^{2}(t)}}\exp\left[-\frac{x^{2}}{2\sigma_{x.\eq}^{2}(t)}\right].
\end{equation}

The situation is more subtle for two-times objects. Indeed, let us consider the same equation but for a given initial condition,
i.e. when we are interested in the transition probability
$p(x,\tau|x_{\ini},0)$, such that
$p(x,\tau|x_{\ini},0)|_{\tau=0}=\delta(x-x_{\ini})$ and
$x(t=0)=x_{\ini}$. Over the time scale $\tau$,
Eq.~\eqref{eq:sigmax-lowest} again holds but it does not verify the
initial condition $\sigma_{x,\ini}=0$. Note that
over the time scale $\tau$,
\begin{equation}\label{eq:phi-tau-scale}
  \varphi(t,0)=\int_{0}^{t}dt' \kappa(t') \quad\rightarrow\quad \varphi(\tau,0)=t_{\fin}\int_{0}^{\tau}d\tau' \kappa(\tau').
\end{equation}
Therefore, the ``external''
approximation Eq.~\eqref{eq:sigmax-lowest}--using the
terminology in Ref.~\cite{bender_advanced_1999}--holds for
$\tau=O(1)$, such that $\varphi=O(t_{\fin})\gg 1$, but not for
short times, such that
$\varphi=O(1)$, where a boundary layer emerges.
In the boundary layer, we obtain the
``internal'' solution
\begin{equation}\label{eq:sigmax-short-times}
  \sigma_{x}^{2}(t)\sim\sigma_{x,\eq}^{2}(t)-
  \sigma_{x,\ini}^{2} e^{-2\varphi(t,0)}.
\end{equation}
This can be justified either by a dominant balance argument in the differential
equation~\eqref{eq:variance-eq-dimless} or by showing that the last
term on the rhs of Eq.~\eqref{eq:chi2-sol-nondim} is subdominant
against the first one. Consistently with the notation employed in the main text, $\sigma_{x,\ini}^2$ stands for the initial value of the variance $\sigma_{x.\eq}^2(t=0)$. A uniform solution in time is obtained by adding
Eq.~\eqref{eq:sigmax-lowest} and \eqref{eq:sigmax-short-times} and
subtracting their common limit for $\varphi(t,0)\gg 1$ and
$\tau\ll 1$, which is $\sigma_{x,\eq}^{2}(t)$. Therefore,
Eq.~\eqref{eq:sigmax-short-times} gives the uniform solution and
$p(x,t|x_{\ini},0)$ is the Gaussian distribution with that variance
and mean $x_{\ini}\exp[-\varphi(t,0)]$, as predicted  by
Eq.~\eqref{eq:average-sol-nondim},
\begin{widetext}
\begin{equation}
  p(x,t|x_{\ini},0)\sim\frac{1}{\sqrt{2\pi\left[ \sigma_{x,\eq}^{2}(t)-\sigma_{x,\ini}^{2}
      e^{-2\varphi(t,0)}\right] }}
\exp\left\{-
    \frac{\left(x-x_{\ini}
        e^{-\varphi(t,0)}\right)^{2}}{2\left[\sigma_{x,\eq}^{2}(t)-
        \sigma_{x,\ini}^2
      e^{-2\varphi(t,0)}\right]}\right\}, \quad t\geq 0.
\end{equation}
This equation can be readily generalised to a given initial condition
$x'$ at time any $t'$ as
\begin{equation}\label{eq:trans-prob}
 p(x,t|x',t')\sim\frac{1}{\sqrt{2\pi\left[ \sigma_{x,\eq}^{2}(t)-\sigma_{x,\eq}^{2}(t')
      e^{-2\varphi(t,t')}\right] }}
\exp\left\{-
    \frac{\left(x-x'
        e^{-\varphi(t,t')}\right)^{2}}{2\left[\sigma_{x,\eq}^{2}(t)
        -\sigma_{x,\eq}^{2}(t')
      e^{-2\varphi(t,t')}\right]}\right\}, \quad t\geq t'.
\end{equation}
\end{widetext}
The lowest order approximation given by
Eqs.~\eqref{eq:one-time-prob} and \eqref{eq:trans-prob} is consistent,
these PDFs obey the Chapman-Kolmogorov conditions $\int dx' p(x,t|x',t')
p(x',t'|x'', t'')=p(x,t|x'',t'')$ and $\int dx' p(x,t|x',t')
p(x',t')=p(x,t)$.

For calculating the probability distributions of energy increment,
work and heat, we will need to calculate correlation functions of the
form
\begin{align}
  C(t,t')\equiv&
  \mean{x^{2}(t)x^{2}(t')}-\mean{x^{2}(t)}\;\mean{x^{2}(t')} \nonumber
  \\
  =&\mean{
    \left(x^{2}(t)-\mean{x^{2}(t)}\right)\left(x^{2}(t')-\mean{x^{2}(t')}
    \right)}.\label{eq:correlations-needed}
\end{align}
In our lowest order approximation, we have that $\mean{x(t)}=0$ and
\begin{equation}
  \mean{x^{2}(t)}\sim\sigma_{x,\eq}^{2}(t)
\end{equation}
for all times. Then, the two-time correlations reduce to
\begin{align}
  C(t,t')  \sim &\mean{
    \left(x^{2}(t)-\sigma_{x,\eq}^{2}(t)\right)
    \left(x^{2}(t')-\sigma_{x,\eq}^{2}(t')
    \right)} \nonumber \\ =&\int dx \int dx' \left(x^{2}-\sigma_{x,\eq}^{2}(t)\right) \nonumber \\ &\times
    \left({x'}^{2}-\sigma_{x,\eq}^{2}(t')\right) p(x,t|x't')p(x',t'). \label{eq:correlations-simpler}
\end{align}
Therefore, we first evaluate the conditioned average
\begin{align}
 & \int dx  \left(x^{2}-\sigma_{x,\eq}^{2}(t)\right)
 p(x,t|x't') \nonumber \\ &=\left({x'}^{2}-\sigma_{x,\eq}^{2}(t')\right)e^{-2\varphi(t,t')},
 \quad t\geq t'. \label{eq:conditioned-average-x2}
\end{align}
which inserted into Eq.~\eqref{eq:correlations-simpler} leads to
\begin{align}
  C(t,t')&\sim e^{-2\varphi(t,t')} \int dx'
  \left({x'}^{2}-\sigma_{x,\eq}^{2}(t')\right)^{2} p(x',t') \nonumber
  \\ &=2
  \sigma_{x,\eq}^{4}(t')e^{-2\varphi(t,t')}, \quad t\geq t'. \label{eq:correlations-result}
\end{align}
Correlations are relevant over the ``fast'' time scale $t$, as long as
$\varphi(t,t')=O(1)$, but become exponentially small over the ``slow''
time scale $\tau$ because, consistently with
Eq.~\eqref{eq:phi-tau-scale}, 
\begin{equation}\label{eq:phi-tau-scale-bis}
  \varphi(t,t')=\int_{t'}^{t} dt'' \kappa(t'') \quad\rightarrow\quad \varphi(\tau,\tau')=t_{\fin}\int_{\tau'}^{\tau}d\tau'' \kappa(\tau''),
\end{equation}
and $\varphi(\tau,\tau')=O(t_{\fin})\gg 1$. In the quasi-static limit
$t_{\fin}\to\infty$, $C(t,t')\to 0$ because $p(x,t|x',t')\to p(x,t)$
and time correlations are ``instantaneously'' killed.

\subsection{Fluctuations of the energy
  increment}\label{sec:energy-inc-fluct}

Since the increment of kinetic energy is for a given protocol fixed and equal to
$\Delta\theta/2$ (in dimensionless variables),
we focus on the fluctuations of the increment of the potential energy
\begin{equation}\label{eq:delta-u}
  \Delta u=\frac{1}{2}\left(\kappa_{\fin}x_{\fin}^{2}-
    \kappa_{\ini}x_{\ini}^{2}\right).
\end{equation}
The average value is straightforward, $\mean{\Delta
  u}=\Delta\theta/2$. Fluctuations are also easy to calculate, since
\begin{equation}\label{eq:delta-u-minus-mean}
  \Delta u-\mean{\Delta u}=\frac{1}{2}\left[\kappa_{\fin}\left(
      x_{\fin}^{2}-\sigma_{x,\fin}^{2}\right)- 
    \kappa_{\ini}\left(x_{\ini}^{2}-\sigma_{x,\ini}^{2}\right)
  \right]
\end{equation}
and the variance is readily written in terms of the correlation function as
\begin{align}
  \sigma_{\Delta u}^{2} &\equiv \mean{\left(\Delta u-
      \mean{\Delta u}\right)^{2}} \nonumber \\&=
  \frac{1}{4}\left[\kappa_{\fin}^{2}C(t_{\fin},t_{\fin})+\kappa_{\ini}^{2}
    C(0,0)-2\kappa_{\fin}\kappa_{\ini}C(t_{\fin},0)\right].
\end{align}
Employing Eq.~\eqref{eq:correlations-result}, we get
\begin{equation}
  \sigma_{\Delta u}^{2}=
  \frac{1}{2}\left[\kappa_{\fin}^{2}\sigma_{x,\fin}^{4}+\kappa_{\ini}^{2}
    \sigma_{x,\ini}^{2}-2\kappa_{\fin}\kappa_{\ini}
    \sigma_{x,\ini}^{4}e^{-2\varphi(t_{\fin},0)}\right].
\end{equation}
For slow driving, the last term on the rhs is exponentially small in
the connecting time $t_{\fin}$, since
\begin{equation}\label{eq:slow-driving-condition}
  e^{-2\varphi(t_{\fin},0)}=e^{-\hat{\kappa} t_{\fin}}, \quad \hat{\kappa}\equiv\int_{0}^{1}d\tau\,\kappa(\tau)=O(1).
\end{equation}
Neglecting this exponentially decreasing term (EDT), we have that
\begin{equation}\label{eq:variance-deltaU}
  \sigma_{\Delta u}^{2}\sim
  \frac{1}{2}\left(\theta_{\fin}^{2}+\theta_{\ini}^{2}\right).
\end{equation}

In conclusion, $\sigma_{\Delta u}$ coincides with that of the
quasi-static limit, except for EDT. In fact, the whole distribution
$\calP(\Delta u)$
\begin{align}
  \calP(\Delta u)=\int dx_{\ini}\int dx_{\fin}\, &\delta\!\left(\Delta
    u-\frac{1}{2}\kappa_{\fin}x_{\fin}^{2}+
    \frac{1}{2}\kappa_{\ini}x_{\ini}^{2}\right) \nonumber \\& \times
  p(x_{\fin},t_{\fin}|x_{\ini},0) p(x_{\ini},0)
\end{align}
coincides with that for the quasi-static limit except for EDT, because
\begin{widetext}
\begin{align}
  p(x_{\fin},t_{\fin}|x_{\ini},0)\sim& \frac{1}{\sqrt{2\pi\left( \sigma_{x,\fin}^{2}-\sigma_{x,\ini}^{2}
      e^{-2\hat{\kappa}t_{\fin}}\right)}}\exp\left[-
    \frac{\left(x-x_{\ini}
        e^{-\hat{\kappa}t_{\fin}}\right)^{2}}{2\left(\sigma_{x,\fin}^{2}
        -\sigma_{x,\ini}^{2}
      e^{-2\hat{\kappa}t_{\fin}}\right)}\right] =p(x_{\fin},t_{\fin})+\text{EDT}.
\end{align}
Then,
\begin{align}
  \calP(\Delta u)\sim\calP_{\qs}(\Delta u)=\int dx_{\ini}\int dx_{\fin} \delta\!\left(\Delta
  u-\frac{1}{2}\kappa_{\fin}x_{\fin}^{2}+
  \frac{1}{2}\kappa_{\ini}x_{\ini}^{2}\right)
  p(x_{\fin},t_{\fin}) p(x_{\ini},0).
\end{align}
This integration has been carried out in
Ref.~\cite{martinez_adiabatic_2015}, 
\begin{equation}
  \calP(\Delta u)=\frac{1}{\pi\sqrt{\theta_{\ini}\theta_{\fin}}}
  \exp\left[-\frac{\Delta\theta}{2\theta_{\ini}\theta_{\fin}}\Delta
    u\right]K_{0}\left(
    \frac{\theta_{\ini}+\theta_{\fin}}
    {2\theta_{\ini}\theta_{\fin}} |\Delta u|
  \right),  
\end{equation}
where $K_{0}$ is the zero-th order modified Bessel function of the
second kind.  The above results are valid for slow--not necessarily
adiabatic--driving between two equilibrium states. In fact,
adiabaticity does not play any role here.

\subsection{Fluctuations of the work}\label{sec:work-fluct}

In dimensionless variables, work is given by
\begin{equation}
  W=\frac{1}{2}\int_{0}^{t_{\fin}}dt\, \frac{d\kappa(t)}{dt} x^{2}(t),
\end{equation}
so that
\begin{align}
  W-\mean{W}=\frac{1}{2}\int_{0}^{t_{\fin}}dt\, \frac{d\kappa(t)}{dt}
  \left(x^{2}(t)-\mean{x^{2}(t)}\right) \sim 
  \frac{1}{2}\int_{0}^{t_{\fin}}dt\, \frac{d\kappa(t)}{dt} 
  \left[x^{2}(t)-\sigma_{x,\eq}^{2}(t)\right]
  .
\end{align}
Therefore, the work variance is
\begin{align}
  \sigma_{W}^{2}\equiv \mean{\left(W-\mean{W}\right)^{2}}=\frac{1}{4}\int_{0}^{t_{\fin}}dt\,\frac{d\kappa(t)}{dt}\int_{0}^{t_{\fin}}dt' \frac{d\kappa(t')}{dt'}
  C(t,t')\nonumber =\frac{1}{2}\int_{0}^{t_{\fin}}dt\,
  \frac{d\kappa(t)}{dt} \int_{0}^{t} dt'\frac{d\kappa(t')}{dt'}
  C(t,t'), \label{eq:sigmaW}
\end{align}
where we have used that $C(t,t')$ and thus the integrand is symmetric
under the exchange $t\leftrightarrow t'$.

Now, we insert Eq.~\eqref{eq:correlations-result} into
\eqref{eq:sigmaW} and go to the slow $\tau$ variable to write
  \begin{align}
  \sigma_{W}^{2}\sim\int_{0}^{1}d\tau\,
 \dot\kappa(\tau) \int_{0}^{\tau} d\tau'
  \dot{\kappa}(\tau')\,
  \sigma_{x,\eq}^{4}(\tau') \exp\left[-2t_{\fin}\int_{\tau'}^{\tau}d\tau''\kappa(\tau'')\right].
\end{align}
For $t_{\fin}\gg 1$, only a narrow region of width $t_{\fin}^{-1}$
contributes to the second integral. Thus, to the
lowest order we can (i) substitute $\tau'$ with $\tau$ in both $\dot{\kappa}(\tau')$ and $\sigma_{x,\eq}(\tau')$
(ii) approximate
$\int_{\tau'}^{\tau}d\tau''\kappa(\tau'')\sim
\kappa(\tau)\Delta$, with $\Delta=\tau-\tau'$, and (iii) extend the  integral over $\Delta$ to the interval $[0,+\infty)$.  Then, we have that 
\begin{equation}
  \int_{0}^{\tau}d\tau'
  \dot{\kappa}(\tau')\sigma_{x,\eq}^4(\tau') \exp\left[-2t_{\fin}\int_{\tau'}^{\tau}d\tau''\kappa(\tau'')\right]
 \sim  \dot{\kappa}(\tau)\sigma_{x,\eq}^4(\tau)\int_{0}^{\infty} d\Delta
 \exp\left[-2t_{\fin}\kappa(\tau)\Delta\right]=
 \frac{\dot\kappa(\tau)\sigma_{x,\eq}^4(\tau)}{2t_{\fin}\kappa(\tau)}.
\end{equation}
and finally the variance for the work reads
\end{widetext}
\begin{align}
  \sigma_{W}^{2}\sim \frac{1}{2t_{\fin}}\int_{0}^{1}d\tau
  \dot{\kappa}^{2}(\tau)
  \frac{\sigma_{x,\eq}^{4}(\tau)}{\kappa(\tau)}
  =\frac{1}{2t_{\fin}}\int_{0}^{1}d\tau
  \dot{\kappa}^{2}(\tau)
  \frac{\theta^{2}(\tau)}{\kappa^{3}(\tau)}. \label{eq:sigmaW-lowest}
\end{align}
Corrections of the order of $t_{\fin}^{-2}$ have been neglected.

In the quasi-static limit $t_{\fin}\to\infty$, the variance vanishes
and work becomes delta-distributed around its mean. For long
$t_{\fin}$ but not infinite, work becomes Gaussian-distributed with
the variance given by Eq.~\eqref{eq:sigmaW-lowest} to the lowest
order. Adiabaticity only enters the picture by restricting the
stiffness and temperature profiles in Eq.~\eqref{eq:sigmaW-lowest}. In
addition, for adiabatic processes $\mean{Q}=0$ and the mean work
$\mean{W}=\Delta E=\Delta\theta$.

\subsection{Fluctuations of the heat}\label{sec:heat-fluct}

We now turn our attention to the fluctuations of the heat. Since the
kinetic contribution is fixed in the overdamped description, this is
equivalent to consider the fluctuations of its configurational contribution
$Q_{x}=\Delta u-W$. Therefore, the deviation from the mean value is
given by
\begin{widetext}
\begin{align}
  Q-\mean{Q}=Q_{x}-\mean{Q_{x}}=\Delta u-\mean{\Delta
u}-\left(W-\mean{W}\right)=\frac{1}{2}\kappa_{\fin}
(x_{\fin}^{2}- \sigma_{x,\fin}^{2}) -\frac{1}{2}\kappa_{\ini}
(x_{\ini}^{2}-\sigma_{x,\ini}^{2})
-\frac{1}{2}\int_{0}^{t_{\fin}}dt \frac{d\kappa(t)}{dt}
[x^{2}(t)-\mean{x^{2}(t)}].
  \label{eq:qx-minus-its-mean}
\end{align}
\end{widetext}
The $n$-th central moment is defined by
\begin{equation}
  \mu_{Q,n}\equiv \mean{\left(Q_{x}-\mean{Q_{x}}\right)^{n}}, \quad
  n\in\mathbb{N}.
\end{equation}
To calculate such moments, we will need to
evaluate $n$-times correlation functions. For the variance, this means
that it suffices to know the two-time correlations introduced in
Eq.~\eqref{eq:correlations-needed}, as has already been the case for
the work fluctuations. More specifically,
\begin{equation}
  \sigma_{Q}^{2}\equiv \mu_{Q,2}=\sigma_{\Delta
  u}^{2}+\sigma_{W}^{2}-2\mean{\left(W-\mean{W}\right)\left(\Delta
    u-\mean{\Delta u}\right)},
\end{equation}
and we focus in the following on the last term, i.e on the calculation of the
energy-work
cross-correlation.

From the definitions of $\Delta u$ and $W$, it is straightforward that
\begin{widetext}
\begin{align}
  \mean{\left(W-\mean{W}\right)\left(\Delta u-\mean{\Delta u}\right)}=
\frac{1}{4}\int_{0}^{t_{\fin}}dt\,\frac{d\kappa(t)}{dt}\left[\kappa_{\fin}C(t_{\fin},t)-\kappa_{\ini}C(t,0)\right],
\end{align}
and making use of Eq.~\eqref{eq:correlations-result} we have that
\begin{align}
  \mean{\left(W-\mean{W}\right)\left(\Delta u-\mean{\Delta
u}\right)}\sim\frac{1}{2} \int_{0}^{t_{\fin}}dt\,\frac{d\kappa(t)}{dt}
\left[\kappa_{\fin}\,\sigma_{x,\eq}^{4}(t)e^{-2\varphi(t_{\fin},t)}-\kappa_{\ini}\sigma_{x,\ini}^{4}e^{-2\varphi(t,0)}\right].
\end{align}
Again, going to the slow variable $\tau$,
\begin{align}
\mean{\left(W-\mean{W}\right)\left(\Delta
    u-\mean{\Delta u}\right)}\sim  \frac{1}{2}
\int_{0}^{1}d\tau\,\dot{\kappa}(\tau) \left\{\kappa_{\fin}\,\sigma_{x,\eq}^{4}(\tau)\exp\left[-2t_{\fin}\int_{\tau}^{1}d\tau'\kappa(\tau')\right]-\kappa_{\ini}\sigma_{x,\ini}^{4}
  \exp\left[-2t_{\fin}\int_{0}^{\tau}d\tau'\kappa(\tau')\right]\right\}.
\end{align}
\end{widetext}
Similarly to the calculation for $\sigma_{W}^{2}$, both terms
contribute in a narrow $\tau$ interval, namely that close to $\tau=1$
($\tau=0$) for the first (second) one. Thus we obtain
\begin{align}
\mean{\left(W-\mean{W}\right)\left(\Delta u-\mean{\Delta
u}\right)}\sim & \frac{1}{4t_{\fin}}
\left[\dot\kappa_{\fin}\,\sigma_{x,\fin}^{4}-\dot\kappa_{\ini}\,\sigma_{x,\ini}^{4}
\right] \nonumber \\ =& \frac{1}{4t_{\fin}} \left[\dot\kappa_{\fin}\,
\frac{\theta_{\fin}^{2}}{\kappa_{\fin}^{2} }-\dot\kappa_{\ini}\,
\frac{\theta_{\ini}^{2}}{\kappa_{\ini}^{2}} \right],
\end{align}
neglecting once more $O(t_{\fin}^{-2})$ corrections. Finally, we get
for the variance of the heat
\begin{align}
  \sigma_{Q}^{2}=\sigma_{\Delta u}^{2}&+\frac{1}{2t_{\fin}}\int_{0}^{1}d\tau
  \left[\dot{\kappa}(\tau)\right]^{2}
  \frac{\theta^{2}(\tau)}{\kappa^{3}(\tau)}
  \nonumber \\&-\frac{1}{2t_{\fin}}
  \left[\dot\kappa_{\fin}
    \frac{\theta_{\fin}^{2}}{\kappa_{\fin}^{2}
    }-\dot\kappa_{\ini}
    \frac{\theta_{\ini}^{2}}{\kappa_{\ini}^{2}}
  \right]+O(t_{\fin}^{-2}).
\end{align}
Integration by parts simplifies this into
\begin{equation}\label{eq:sigma!-result}
  \sigma_{Q}^{2}=
  \sigma_{\Delta u}^{2}-\frac{1}{2t_{\fin}}
  \int_{0}^{1}d\tau\,\kappa(\tau)\frac{d}{d\tau}
  \left[
    \dot\kappa(\tau) \frac{\theta^{2}(\tau)}{\kappa^{3}(\tau)}
  \right]
  +O(t_{\fin}^{-2}).
\end{equation}
This expression is also valid for slow driving, regardless of being
adiabatic or not. Similarly to the case of the work distribution,
adiabaticity only enters the picture by restricting the stiffness and
temperature profiles that can be substituted into Eq.~\eqref{eq:sigma!-result}.

Therefore, in the limit of slow driving we find a small change in the
variance of the heat--recall that it is non-zero and equal to
$\sigma_{\Delta u}^{2}$ for the quasi-static case. For slow but not
quasi-static driving, work is no longer delta-distributed around the
mean and then the fluctuations of heat and energy increment  are not
equivalent: the corrections are of the order of $t_{\fin}^{-1}$ for
the former but exponentially small for the latter. A relevant question
thus arises: whether or not the heat distribution conserves its shape,
i.e if all the change of the distribution can be encoded in the change
of the variance. Although the calculation of the whole heat
distribution seems to be a challenging mathematical problem--even for the harmonic case, we show that the situation is more complex
in the following, by obtaining the third central moment of the
distribution--recall that the distribution of the heat is asymmetric
around its mean.

Consistently with the comments above, we consider the third central
moment $\mu_{Q,3}$. Making use of Eq.~\eqref{eq:qx-minus-its-mean},
we have that
\begin{align}
  \mu_{Q,3}=&\mean{(\Delta u-\mean{\Delta u})^{3}}-3\mean{(\Delta u-\mean{\Delta u})^{2}(W-\mean{W})}\nonumber \\&+3\mean{(\Delta u-\mean{\Delta u})(W-\mean{W})^{2}}-\mean{(W-\mean{W})^{3}}.
\end{align}
In order to obtain $\mu_{Q,3}$, we need to evaluate
three-time correlations of the kind
\begin{widetext}
\begin{equation}
A(t_{1},t_{2},t_{3})=\mean{\left[x^{2}(t_{1})-\mean{x^{2}(t_{1})}\right]\left[x^{2}(t_{2})-\mean{x^{2}(t_{2})}\right]\left[x^{2}(t_{3})-\mean{x^{2}(t_{3})}\right]},
  \quad t_{1}\geq t_{2}\geq t_{3}.
\end{equation}
\end{widetext}
In the same approximation employed throughout these appendices, i.e. that given by Eqs.~\eqref{eq:one-time-prob} and
\eqref{eq:trans-prob}, this correlation has the asymptotic behaviour
\begin{equation}
  A(t_{1},t_{2},t_{3})\sim 8
  \sigma_{x,\eq}^{2}(t_{2})\sigma_{x,eq}^{4}(t_{3})\exp[-2\varphi(t_{1},t_{3})].
\end{equation}
In the following, we repeatedly use this expression for
$A(t_{1},t_{2},t_{3})$ to calculate all  contributions to
$\mu_{Q,3}$.

We start by considering
\begin{equation}
  \mu_{Q,3}^{(1)}\equiv \mean{(\Delta u-
    \mean{\Delta u})^{3}}=
  \mean{ \left[\frac{1}{2}\kappa_{\fin}(x_{\fin}^{2}-\sigma_{x,\fin}^{2})
    -\frac{1}{2}\kappa_{\ini}(x_{\ini}^{2}-\sigma_{x,\ini}^{2})\right]^{3} }.
\end{equation}
In this case, there is no integration and any term that mixes
$(x_{\fin}^{2}-\sigma_{x,\fin}^{2})$ and
$(x_{\ini}^{2}-\sigma_{x,\ini}^{2})$ contains an EDT of the form
$\exp[-2\varphi(t_{\fin},0)]=\exp[-2\hat{\kappa}t_{\fin}]$. Thus, we recover the quasi-static situation in which $(x_{\fin}^{2}-\sigma_{x,\fin}^{2})$ and
$(x_{\ini}^{2}-\sigma_{x,\ini}^{2})$ are uncorrelated, except
for EDT, i.e.
\begin{equation}
  \mu_{Q,3}^{(1)}\sim\mu_{Q,3}^{\qs}=\frac{1}{8}\kappa_{\fin}^{3}
  \mean{(x_{\fin}^{2}-\sigma_{x,\fin}^{2})^{3}}
  -\frac{1}{8}\kappa_{\ini}\mean{(x_{\ini}^{2}-\sigma_{x,\ini}^{2})^{3}}=\theta_{\fin}^{3}-\theta_{\ini}^{3}.
\end{equation}
If $\theta_{\fin}\neq\theta_{\ini}$, this contribution is different
from zero, in accordance with the heat fluctuations being asymmetric
around its mean in the quasi-static limit.

Now, we turn our attention to
\begin{widetext}
\begin{equation}
  \mu_{Q,3}^{(2)}\equiv -3\mean{(\Delta u-\mean{\Delta
      u})^{2}(W-\mean{W})}\sim -\frac{3}{2}
  \mean{\left[\frac{1}{2}\kappa_{\fin}(x_{\fin}^{2}-\sigma_{x,\fin}^{2})
      -\frac{1}{2}\kappa_{\ini}(x_{\ini}^{2}-\sigma_{x,\ini}^{2})\right]^{2}
    \int_{0}^{t_{\fin}}dt\, \frac{d\kappa(t)}{dt}
    \left[x^{2}(t)-\sigma_{x,\eq}^{2}(t)\right] } ,
\end{equation}
i.e.
\begin{align}
   \mu_{Q,3}^{(2)} \sim -\frac{3}{8}\int_{0}^{t_{\fin}}dt\,
  \frac{d\kappa(t)}{dt} \left[\kappa_{\fin}^{2} A(t_{\fin},t_{\fin},t)-2\kappa_{\fin}\kappa_{\ini}A(t_{\fin},t,0)+\kappa_{\ini}^{2}A(t,0,0)\right].
\end{align}
The term containing the correlation $A(t_{\fin},t,0)$ is exponentially
decreasing,  $A(t_{\fin},t,0)\sim
8\sigma_{x,\eq}^{2}(t)\sigma_{x,\ini}^{4}\exp[-2\varphi(t_{\fin},0)]$. Then,
we only need to consider the other two terms. We start with the analysis of the first one, specifically
\begin{align}
  \int_{0}^{t_{\fin}}dt\, \frac{d\kappa(t)}{dt} A(t_{\fin},t_{\fin},t)
\sim
8\int_{0}^{t_{\fin}}dt\,\frac{d\kappa(t)}{dt}\sigma_{x,\fin}^{2}\,\sigma_{x,\eq}^{4}(t)\exp[-2\varphi(t_{\fin},t)]
=8\sigma_{x,\fin}^{2}\int_{0}^{1} d\tau
\dot\kappa(\tau)\sigma_{x,\eq}^{4}(\tau)
\exp\left[-2t_{\fin}\int_{\tau}^{1}d\tau' \kappa(\tau')\right].
\end{align}
In the limit $t_{\fin}\gg 1$, we can once more use Watson's lemma to
estimate the integral to the lowest order, with the result
\begin{equation}
  \int_{0}^{t_{\fin}}dt\,
  \frac{d\kappa(t)}{dt}
    A(t_{\fin},t_{\fin},t)\sim
    8\sigma_{x,\fin}^{6}\dot\kappa_{\fin}
    \frac{1}{2\kappa_{\fin}t_{\fin}}=
    \frac{1}{t_{\fin}}
    \frac{4\sigma_{x,\fin}^{6}}{\kappa_{\fin}}
    \dot\kappa_{\fin}.
\end{equation}
An analogous calculation yields
\begin{equation}
  \int_{0}^{t_{\fin}}dt\,
  \frac{d\kappa(t)}{dt}
    A(t,0,0)\sim
    8\sigma_{x,\ini}^{6}\dot\kappa_{\ini}
    \frac{1}{2\kappa_{\ini}t_{\fin}}=
    \frac{1}{t_{\fin}}\frac{4\sigma_{x,\ini}^{6}}{\kappa_{\ini}}
    \dot\kappa_{\ini}.
\end{equation}
Then, up to order $t_{\fin}^{-1}$ we have that
\begin{equation}
  \mu_{Q,3}^{(2)} \sim
  -\frac{3}{2t_{\fin}}\left(\dot\kappa_{\fin}\frac{\theta_{\fin}^{3}}{\kappa_{\fin}^2}-\dot\kappa_{\ini}\frac{\theta_{\ini}^{3}}{\kappa_{\ini}^2}\right).
\end{equation}

Let us analyse the following contribution
\begin{align}
  \mu_{Q,3}^{(3)}\equiv & 3\mean{(\Delta u-\mean{\Delta
      u})(W-\mean{W})^2} \nonumber \\
      = & \frac{3}{4}
  \mean{\left[\frac{1}{2}\kappa_{\fin}(x_{\fin}^{2}-\sigma_{x,\fin}^{2})
      -\frac{1}{2}\kappa_{\ini}(x_{\ini}^{2}-\sigma_{x,\ini}^{2})\right]
    \int_{0}^{t_{\fin}}dt\, \frac{d\kappa(t)}{dt}
    \left[x^{2}(t)-\mean{x^2(t)}\right]
    \int_{0}^{t_{\fin}}dt'\, \frac{d\kappa(t')}{dt'}
    \left[x^{2}(t')-\mean{x^2(t')}\right]
    }
     \nonumber \\
    = & \frac{3}{2}
  \mean{\left[\frac{1}{2}\kappa_{\fin}(x_{\fin}^{2}-\sigma_{x,\fin}^{2})
      -\frac{1}{2}\kappa_{\ini}(x_{\ini}^{2}-\sigma_{x,\ini}^{2})\right]
    \int_{0}^{t_{\fin}}dt\, \frac{d\kappa(t)}{dt}
    \left[x^{2}(t)-\mean{x^2(t)}\right]
    \int_{0}^{t}dt'\, \frac{d\kappa(t')}{dt'}
    \left[x^{2}(t')-\mean{x^2(t')}\right]}
    \nonumber \\
    = & \frac{3}{4} \int_0^{t_{\fin}} dt \frac{d\kappa(t)}{dt} \int_0^t dt' \frac{d\kappa(t')}{dt'}\left[\kappa_{\fin} A(t_{\fin},t,t')-\kappa_{\ini} A(t,t',0)\right]
    .
\end{align}
Note that $t\geq t'$ in the last two lines, making use of the symmetry of the integrand under the exchange $t\leftrightarrow t'$.

Again, we have to calculate three-times correlation functions. We start by analysing the term stemming from
\begin{equation}
    A(t_{\fin},t,t')\sim 8 \sigma_{x,\eq}^2(t)\sigma_{x,\eq}^4(t')\exp[-2\varphi(t_{\fin},t')].
\end{equation}
More specifically, we have to find the lowest order contribution to the integral
\begin{align}
    \int_{0}^{t_{\fin}} dt \frac{d\kappa(t)}{dt} \int_0^{t} dt' \frac{d\kappa(t')}{dt'}
    A(t_{\fin},t,t')\sim & 8 \int_{0}^{t_{\fin}} dt \frac{d\kappa(t)}{dt} \int_0^{t} dt' \frac{d\kappa(t')}{dt'}
    \sigma_{x,\eq}^2(t)\sigma_{x,\eq}^4(t') \exp[-2\varphi(t_{\fin},t')] \nonumber \\
    = & 8 \int_{0}^1 d\tau \dot\kappa(\tau) \sigma_{x,\eq}^2(\tau)
    \int_0^{\tau} d\tau' \dot\kappa(\tau') \sigma_{x,\eq}^4(\tau')\exp{\left[-2t_{\fin}\int_{\tau'}^1 d\tau'' \kappa(\tau'') \right]}.
\end{align}
Once more, the asymptotic estimate of this integral to the lowest order can be calculated by applying Watson's lemma. First, we integrate over $\tau'$ at given $\tau$, and this yields    
\begin{equation}
    \int_{0}^{t_{\fin}} dt \frac{d\kappa(t)}{dt} \int_0^{t} dt' \frac{d\kappa(t')}{dt'}
    A(t_{\fin},t,t')\sim \int_{0}^1 d\tau \dot{\kappa}^{2}(\tau) \sigma_{x,\eq}^6(\tau)
    \frac{1}{2\kappa(\tau)t_{\fin}}\exp{\left[-2t_{\fin}\int_{\tau}^1 d\tau'' \kappa(\tau'') \right]},
\end{equation}
\end{widetext}
because the exponential reaches its maximum value for $\tau'=\tau$, i.e. when $\tau'$ is closest to unity. The integral over $\tau$ is now dominated by the contribution of a narrow interval close to $\tau=1$, i.e by applying again Watson's lemma we get
\begin{equation}
    \int_{0}^{t_{\fin}} dt \frac{d\kappa(t)}{dt} \int_0^{t} dt' \frac{d\kappa(t')}{dt'}
    A(t_{\fin},t,t')\sim \dot\kappa_{\fin}^2 \sigma_{x,\fin}^6
    \left(\frac{1}{2\kappa_{\fin}t_{\fin}}\right)^2.
\end{equation}
Therefore, this contribution is of the order $t_{\fin}^{-2}$ and thus subdominant to that in $\mu_{Q,2}^{(2)}$, which was of the order of $t_{\fin}^{-1}$.

Now we look into the contribution coming from $A(t,t',0)$. In this
case, it is better to introduce the condition $t\geq t'$ by
integrating $t'$ from $0$ to $t_{\fin}$ and restricting $t$ to the
interval $[t',t_{\fin}]$. By doing so, the calculation follows
completely similar lines as those above. The correlation $A(t,t',0)$
has a term $\exp[-2\varphi(t,0)]=\exp[-2t_{\fin}\int_0^{\tau} d\tau''
\kappa(\tau'')]$: the first integration over $t$ gives a
$t_{\fin}^{-1}$ factor and makes $t=t'$ in $\varphi$, the second
integration over $t'$ gives a second $t_{\fin}^-1$ factor and makes
$t'=0$ in $\varphi$. Then, we have that this contribution is also proportional to $t_{\fin}^{-2}$, specifically
\begin{equation}
    \int_{0}^{t_{\fin}} dt \frac{d\kappa(t)}{dt} \int_0^{t} dt' \frac{d\kappa(t')}{dt'}
    A(t,t',0)\sim \dot\kappa_{\ini}^2 \sigma_{x,\ini}^6
    \left(\frac{1}{2\kappa_{\ini}t_{\fin}}\right)^2.
\end{equation}

With a similar line of reasoning, it is possible to show that the last  contribution to $\mu_{Q,3}$,
\begin{equation}
    \mu_{Q,3}^{(4)}=-\mean{(W-\mean{W})^3},
\end{equation}
is also subdominant--i.e. it does not contain $t_{\fin}^{-1}$ terms. This can also be qualitatively understood by recalling that work fluctuations are Gaussian in the slow driving limit $t_{\fin}\gg 1$: $\mean{(W-\mean{W})^3}$ should vanish to the lowest order. 

Finally, we get that
\begin{equation}
    \mu_{Q,3}=\theta_{\fin}^{3}-\theta_{\ini}^{3}-\frac{3}{2t_{\fin}}\left(\dot\kappa_{\fin}\frac{\theta_{\fin}^{3}}{\kappa_{\fin}^2}-\dot\kappa_{\ini}\frac{\theta_{\ini}^{3}}{\kappa_{\ini}^2}\right)+O(t_{\fin}^{-2}),
\end{equation}
and the correction to the third central moment is a pure boundary term. Interestingly, this means that the heat distribution is not simply being compressed/decompressed around its mean. The heat PDF $\calP(Q)$ is such that
\begin{equation}
    \calP(Q)\neq f\left(\frac{Q-\mean{Q}}{\sigma_{Q}}\right),
\end{equation}
even in the slow driving limit. Had we $\calP(Q)= f\left(\frac{Q-\mean{Q}}{\sigma_{Q}}\right)$, the third central moment would be proportional to $\sigma_{Q}^3$. In other words, $\mu_{Q,3}/\sigma_{Q}^3$ would be constant, independent of $t_{\fin}$, to the considered order. It is quite clear that $\mu_{Q,3}/\sigma_{Q}^3$ does depend on $t_{\fin}$, i.e. the $t_{\fin}^{-1}$ corrections coming from $\mu_{Q,3}$ and $\sigma_{Q}^3$ do not cancel out. Thus, the shape of the heat distribution is not preserved when we change the connecting time.
}

\begin{acknowledgments}
  This work has been financially supported by UNIPD STARS Stg (CdA
  Rep.  40, 23.02.2018) BioReACT grant (C.A.P.), the Agence Nationale
  de la Recherche research funding Grant No.~ANR-18-CE30-0013
  (D.G.-O., E.T.), and by the Spanish Ministerio de Ciencia,
  Innovaci\'on y Universidades through Grant (partially financed by
  the ERDF) No.~PGC2018-093998-B-I00 (A.P.).
\end{acknowledgments}

\bibliography{Mi-biblioteca-21-feb-2020}

\end{document}